\documentclass[11pt,a4paper]{article}

\usepackage{jcappub, natbib}

\allowdisplaybreaks

\usepackage{ifthen}
\usepackage{graphicx}
\usepackage{dcolumn}
\usepackage{bm}
\usepackage{color}
\usepackage{amsmath}
\usepackage{amssymb}
\usepackage{bbold}

\newcommand{\hMs}{\ h^{-1} M_\odot}
\newcommand{\mpc}{\, {\rm Mpc}}

\newcommand{\erg}{\, {\rm erg}}
\newcommand{\seg}{\, {\rm s}}
\newcommand{\hz}{\, {\rm Hz}}
\newcommand{\hmpc}{\, h^{-1} \mpc}

\newcommand{\kms}{\, {\rm km\, s}^{-1}}

\newcommand{\lya}{Ly$\alpha$\ }

\newcommand{\cm}{{\rm cm}}

\begin{document}

\title{The large-scale Quasar-Lyman $\alpha$ Forest Cross-Correlation from BOSS}
\author[a,b]{Andreu Font-Ribera}
\author[c]{, Eduard Arnau}
\author[d,c]{, Jordi Miralda-Escud\'{e}}
\author[e]{, Emmanuel Rollinde}
\author[f]{, J. Brinkmann}
\author[g]{, Joel R. Brownstein}
\author[h]{, Khee-Gan Lee}
\author[i]{, Adam D. Myers}
\author[j]{, Nathalie Palanque-Delabrouille}
\author[k]{, Isabelle P\^aris}
\author[e]{, Patrick Petitjean}
\author[j]{, James Rich}
\author[b]{, Nicholas P. Ross}
\author[l,m]{, Donald P. Schneider}
\author[b,n]{and Martin White}

\affiliation[a]{Institute of Theoretical Physics, University of Zurich,
        8057 Zurich, Switzerland}
\affiliation[b]{Lawrence Berkeley National Laboratory,
	1 Cyclotron Road, Berkeley, CA, USA}
\affiliation[c]{Institut de Ci\`{e}ncies del Cosmos (IEEC/UB),
        Barcelona, Catalonia}
\affiliation[d]{Instituci\'o Catalana de Recerca i Estudis Avan\c cats,
        Catalonia}
\affiliation[e]{Universit\'e Paris 6 et CNRS, Institut d'Astrophysique
        de Paris, 98bis blvd. Arago, 75014 Paris, France}
\affiliation[f]{Apache Point Observatory, P.O. Box 59, Sunspot, NM 88349, USA}
\affiliation[g]{Department of Physics and Astronomy, University of Utah, 
	115 S 1400 E, Salt Lake City, UT 84112, USA}
\affiliation[h]{Max-Planck-Institut f\"ur Astronomie, K\"onigstuhl 17, D-69117
        Heidelberg, Germany}
\affiliation[i]{Department of Physics and Astronomy, University of Wyoming, 
	Laramie, WY 82071, USA}
\affiliation[j]{CEA, Centre de Saclay, IRFU, 91191 Gif-sur-Yvette, France}
\affiliation[k]{Departamento de Astronom\'ia, Universidad de Chile, 
	Casilla 36-D, Santiago, Chile}
\affiliation[l]{Department of Astronomy and Astrophysics, 
	The Pennsylvania State University, University Park, PA 16802, USA}
\affiliation[m]{Institute for Gravitation and the Cosmos, 
	The Pennsylvania State University, University Park, PA 16802, USA}
\affiliation[n]{Departments of Physics and Astronomy, 601 Campbell Hall,
        University of California Berkeley, CA 94720, USA}

\emailAdd{font@physik.uzh.ch}

\abstract{
We measure the large-scale cross-correlation of quasars with the
\lya forest absorption in redshift space, using $\sim 60000$ quasar
spectra from Data Release 9
(DR9) of the Baryon Oscillation Spectroscopic Survey (BOSS).
The cross-correlation is detected over a wide range of scales, up to
comoving separations $r$ of $80$ $\hmpc$. For $r > 15 \hmpc$, we show
that the cross-correlation is well fitted by the linear theory
prediction for the mean overdensity around a quasar host halo in the
standard $\, \rm\Lambda$CDM model, with the redshift distortions indicative
of gravitational evolution detected at high confidence.
Using previous determinations of the \lya forest bias factor obtained
from the \lya autocorrelation, we infer the quasar bias factor to be
$b_q = 3.64^{+0.13}_{-0.15}$ at a mean redshift $z=2.38$, in agreement
with previous measurements from the quasar auto-correlation. We also
obtain a new estimate of the \lya forest redshift distortion factor,
$\beta_F = 1.1 \pm 0.15$, slightly larger than but consistent with the
previous measurement from the \lya forest autocorrelation.
The simple linear model we use fails at separations $r < 15 \hmpc$,
and we show that this may reasonably be due to the enhanced ionization
due to radiation from the quasars.
We also provide the expected correction that the mass overdensity around 
the quasar implies for measurements of the ionizing radiation background 
from the line-of-sight proximity effect.
}


\maketitle

\section{Introduction}

  As the most optically luminous objects known in the universe, quasars
are used as lampposts at high redshift to obtain
absorption spectra of the intervening intergalactic medium, as well as
tracers of large-scale structure. Their absorption spectra blueward of
the \lya emission line reveal the \lya forest, reflecting the structure
in the hydrogen gas density in the intergalactic medium as it evolves
through gravitational collapse around dark matter halos in which
galaxies form (e.g., \cite{1971ApJ...164L..73L},\cite{1998ARA&A..36..267R}, 
\cite{2006ApJS..163...80M}, \cite{2011JCAP...09..001S}).

The large-scale clustering of quasars was measured in the 2dF survey 
(e.g. \cite{PMN04}, \cite{2005MNRAS.356..415C}, \cite{2005MNRAS.360.1040D})
and in the Sloan Digital Sky Survey (SDSS, \cite{2006ApJ...638..622M},
\cite{2007AJ....133.2222S}, \cite{2009ApJ...697.1634R}).
Both the \lya forest and the quasar clustering can be used as tracers
of the underlying large-scale mass fluctuations, which are thought to
have an origin in the initial conditions of the universe.
In the linear regime, the observed quasar correlation function should be
equal to the mass autocorrelation times the square of the mean bias
factor of the quasar host halos.
Recent results from the analyses of data from large-scale 
surveys have indicated a bias factor that increases with redshift and
has a value $b_q = 3.8 \pm 0.3$ at $z=2.4$, and is nearly independent of
quasar luminosity (see \cite{2009ApJ...697.1656S},
\cite{2012MNRAS.424..933W}).

%

  Quasar clustering can also be probed by means of the
cross-correlation with other tracers. The quasar cross-correlation with
galaxies was measured by \cite{2005ApJ...627L...1A},
\cite{2007ApJ...654..115C} and \cite{2009MNRAS.397.1862P}. These studies
found that the quasar bias factor has a value near unity, comparable to
typical star-forming galaxies, at redshift $z \lesssim 1$, but the small
samples at higher redshift already indicated a larger clustering
amplitude.
Quasars can also be
cross-correlated with absorption systems found in the spectra of other
quasars. This can be done with the hydrogen \lya forest, with a high
abundance of absorption features, and also with more sparse metal
line systems such as the CIV lines, which was recently accomplished by
\cite{Vikas2013}.

  The cross-correlation of quasars with the \lya forest absorption was
first searched for along the same line of sight of each individual
quasar, looking for the impact of the quasar ionizing radiation reducing
the \lya absorption, which has been designated as ``proximity effect'' or
``inverse effect'' (\cite{1982MNRAS.198...91C}, 
\cite{1986ApJ...309...19M}, \cite{1988ApJ...327..570B}).
The ionizing radiation emitted by a
quasar is added to the intensity of the cosmic ionizing background, and
the higher than average intensity in the quasar vicinity implies an
increased degree of ionization of the intergalactic medium, and
therefore a decreased absorption by the \lya forest. Studies of the
proximity effect have therefore expected a lower than average
absorption of the \lya forest near the quasar \lya emission line
compared to the absorption at the same redshift seen in quasars at
higher redshift, using this to infer the intensity of the cosmic
ionizing background (see \cite{2000ApJS..130...67S} for a more recent study).
These investigations have generally not included in the analysis the fact that
the intergalactic gas should have higher density near a quasar, because
of the positive correlation of the quasar host halo with the mass density.
In reality, the observations of the quasar-\lya absorption
cross-correlation should reveal the combined effect of the mean
overdensity and the additional ionizing intensity around quasars, which
substantially complicates the theoretical interpretation.

  In general, the quasar-\lya cross-correlation can be measured in
redshift space from \lya forest lines of sight near another quasar,
as a function of the perpendicular and parallel components of the
separation, $\sigma$ and $\pi$, respectively. The effect of a quasar
on the \lya forest in the spectrum of another nearby quasar was
investigated in several papers that examined individual quasar
pairs, generally separated by a few arc minutes (\cite{1986ApJ...303L..27J},
\cite{1989ApJ...336..550C}, \cite{1992A&A...258..234M}, 
\cite{2001MNRAS.328..653L}). These
observations generally found no evidence for any decrease of \lya
absorption near quasars due to the excess ionizing radiation. In fact,
both in the case of quasar pairs at small separation, and in the case
of using a larger number of pairs at wider separations, it
has been found that the \lya absorption is stronger near quasars,
rather than weaker (\cite{1998ApJ...500..525S}, 
\cite{2004ApJ...610..105S}, \cite{2004ApJ...610..642C},
\cite{2005MNRAS.361.1015R},\cite{2007MNRAS.377..657G}),
\cite{2008MNRAS.387..377K}.
This result has been attributed to the mean overdensity near a quasar, 
combined with 
the reduction of the ionizing intensity from the quasar due to both
anisotropic emission and time variability of the quasars, as discussed
by \cite{1992A&A...258..234M}, \cite{2004ApJ...610..105S} and 
\cite{2004ApJ...610..642C}. 
These works obtained upper limits to the luminosity of 
the quasars emitted in the perpendicular direction with a time delay, 
although no detailed analysis was done to attempt to model the effect of the
overdensity and provide a robust interpretation of the data.



  This question can now be investigated with the large sample of quasar
spectra provided by the Baryon Oscillation Spectroscopic Survey (BOSS) of the 
SDSS-III Collaboration (\cite{2011AJ....142...72E}, 
\cite{2013AJ....145...10D}). The DR9 Catalogue of quasars
(\cite{2012A&A...548A..66P}) already contains 87822 quasars,
with more than 60000 of them at $z>2$,
distributed over 3275 square degrees of the sky. The extensive area covered
and the large number of quasars makes this sample particularly useful to study
the quasar-\lya cross-correlation at large scales (i.e., the typical
nearest neighbor projected separation of $\sim 10 \hmpc$ and larger). 

At these separations, the cross-correlation function induced by the 
mass overdensity around quasars can be predicted from linear theory,
and should be proportional to the product of the bias factors of the
quasars and the \lya forest and show the expected redshift distortions
(\cite{1987MNRAS.227....1K}, \cite{HAMIL92}). These bias factors can be
independently determined by observations of the autocorrelations of the 
\lya forest and the quasars (\cite{2011JCAP...09..001S}, 
\cite{2012MNRAS.424..933W}), and should agree with the measured amplitude
of the cross-correlation.

  This paper presents the quasar-\lya cross correlation measured from
the quasars in the DR9 catalogue of BOSS. The data sample is
presented in section \ref{sec:dataset}, and the method of analysis in
section \ref{sec:method}. Linear
theory is used to model the mean cross-correlation, and the fits to the
observational results are presented in section \ref{sec:results}, 
showing that the overdensity effect alone adequately matches the results 
at comoving separations $r > 15 \hmpc$ for reasonable values of the 
bias factors. Section \ref{sec:discussion} discusses the expected effect 
of the ionizing radiation of the quasars on the cross-correlation for 
the quasar luminosities in our sample, and the manner that this may impact the 
results in section \ref{sec:results}. Our conclusions are 
summarized in section \ref{sec:conclusions}.

Throughout this paper we use the flat $\, \rm\Lambda$CDM cosmology, with
$\Omega_m = 0.281$, $\Omega_b = 0.0462$, $h=0.71$, $n_s=0.963$ and
$\sigma_8=0.8$, consistent with the cosmological parameters from the WMAP
mission \cite{2011ApJS..192...18K}.

\section{Data Set}
\label{sec:dataset}

  The data used in this paper is from the publicly available 9th
Data Release (DR9, \cite{2012ApJS..203...21A})
of the SDSS-III Collaboration 
(\cite{2011AJ....142...72E}, 
\cite{2012AJ....144..144B}, 
\cite{1998AJ....116.3040G}, 
\cite{1996AJ....111.1748F}, 
\cite{2006AJ....131.2332G}, 
\cite{2012arXiv1208.2233S}, 
\cite{2000AJ....120.1579Y}), 
comprising the first two
years of observations of the Baryon Oscillation Spectroscopic Survey
(BOSS, \cite{2013AJ....145...10D}). 
The quasar sample is a subsample of the catalogue
described in \cite{2012A&A...548A..66P}, while the \lya absorption sightlines
make use of the products described in \cite{2013AJ....145...69L}. 

We first describe the cuts we apply to the catalogue to select our
quasar sample, and then the set of quasar lines of sight that we use
for the \lya absorption field to be cross-correlated with the quasars.

\subsection{Quasar sample}

A total of 87822 quasars are present in the DR9 quasar catalogue
(\cite{2012A&A...548A..66P}). These quasars were targeted for spectroscopy
using a complex target selection procedure presented in 
\cite{2012ApJS..199....3R} 
that combines a series of methods described in 
\cite{2010A&A...523A..14Y}, 
\cite{2011ApJ...743..125K}, 
and \cite{2011ApJ...729..141B}. 
The French Participation Group (FPG) of the SDSS-III Collaboration verified 
each of these objects by visually inspecting the spectra.
A number of estimates of the quasar redshift based on different methods
are provided in this catalogue. We generally use the redshift 
obtained using the Principal Component Analysis method (Z\_PCA in DR9Q), 
but in Section \ref{ss:redshifts} we examine our results for the quasar-\lya
cross-correlation when using other estimates for the quasar redshifts.

\begin{figure}
 \begin{center}
  \includegraphics[scale=0.5, angle=-90]{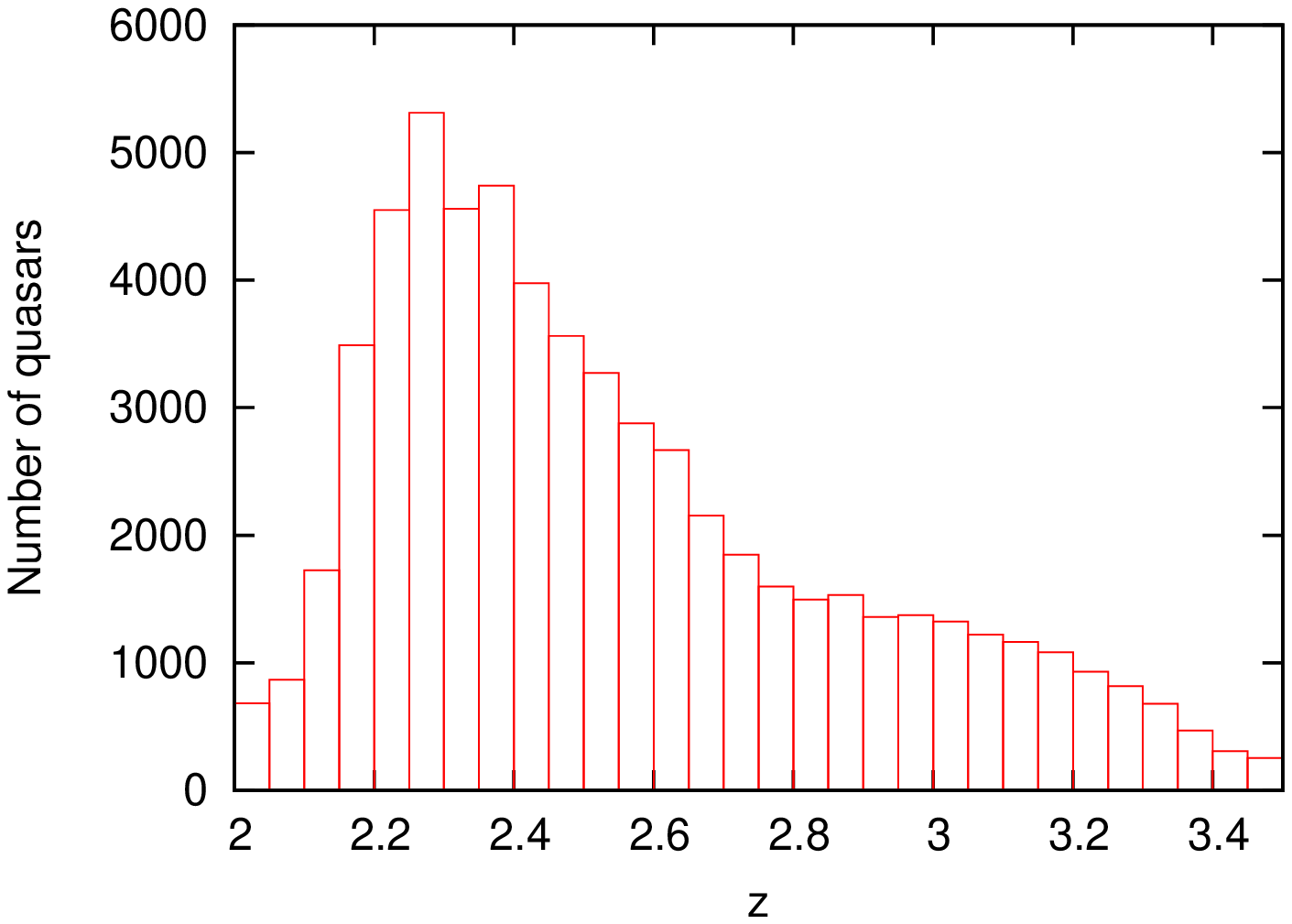}
  \includegraphics[scale=0.5, angle=-90]{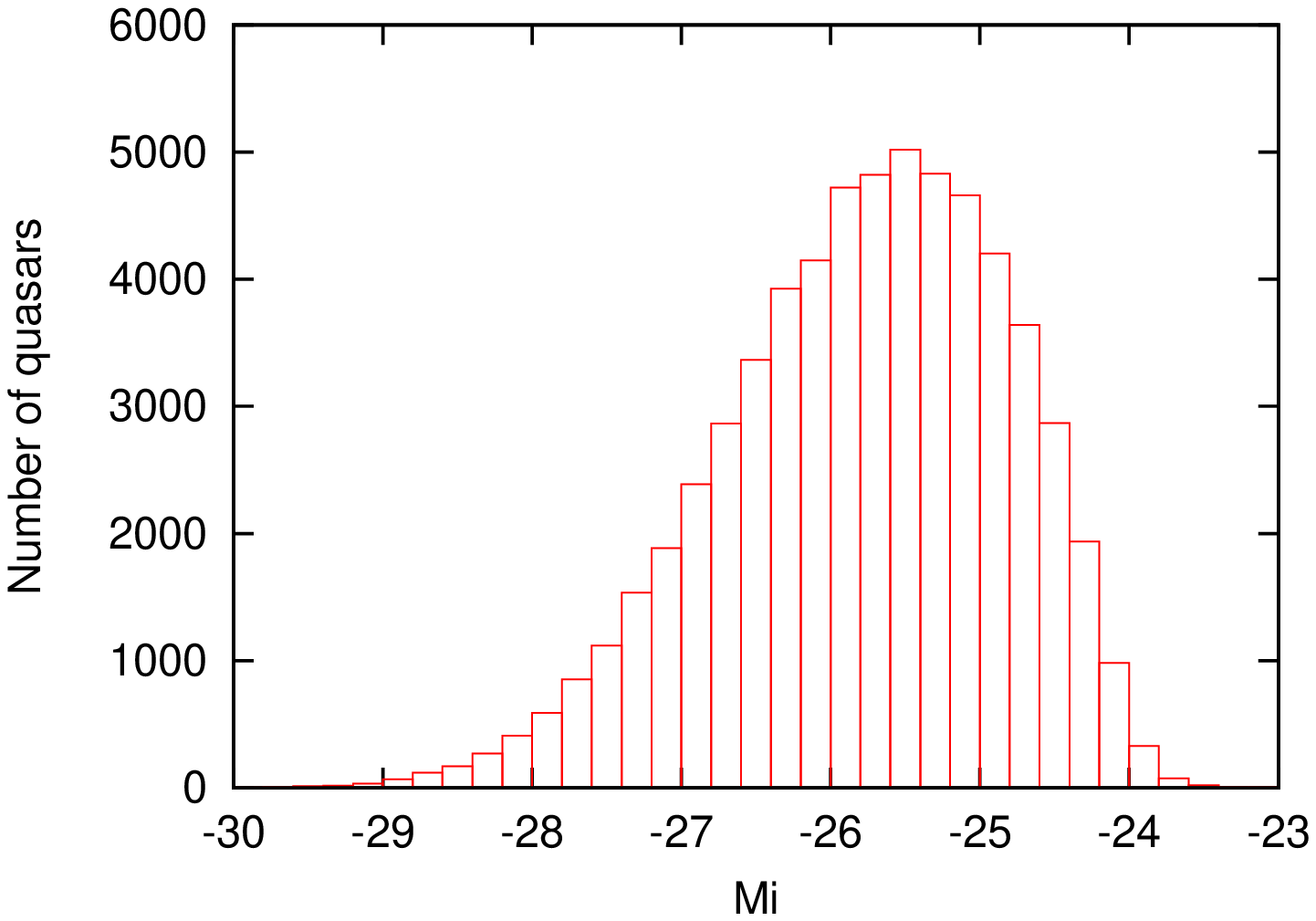}
 \end{center}
 \caption{Left panel: Distribution of the 61342 quasar redshifts in our
sample. Right panel: $i$-band absolute magnitude distribution.}
 \label{fig:dndz}
\end{figure}

In order to have a well defined redshift interval in our sample, we use
only the 61366 quasars with a redshift in the range $2 < z_q < 3.5$. 
Most of the 26456 quasars discarded by this criteria are at a redshift
that is too low to be cross-correlated with the observed \lya absorption
spectra, and the few at $z_q>3.5$ have too few high-redshift nearby
lines of sight to yield useful results. Finally,
we apply a cut in $i$-band absolute magnitude, $-30 < M_i < -23$, which
reduces the final sample to 61342 quasars. The magnitudes are provided in the 
quasar catalog (\cite{2012A&A...548A..66P}), and were computed using a 
similar cosmological model than the one used in this study.
The redshift and $i$-band absolute magnitude distributions of our quasar
sample are shown in figure \ref{fig:dndz}.

\subsection{\lya sample}

Not all quasars present in the catalogue are useful and free of
systematic effects for using the \lya forest spectrum: they may be
affected by broad absorption lines, they may have too low a redshift
so that their \lya forest pixels lie in the noisy blue end of 
the BOSS spectrograph, or their continuum may be too difficult to model.
For these reasons, we select the same quasars as in 
{\cite{2012JCAP...11..059F}, a previous study of the cross-correlation 
of the \lya forest with Damped \lya Systems (DLAs) using the DLA 
catalogue of \cite{2012A&A...547L...1N}.
This reduces the number of available lines of sight to $52449$.
The selection criteria are that the quasar redshift lies in the range
$2.15 < z < 3.5$, that the BAL\_FLAG\_VI flag is not set in the catalogue of
\cite{2012A&A...548A..66P}, and that the median signal-to-noise ratio
per pixel (of width $\sim 69 \kms$) in the quasar rest-frame wavelength range
$1220 {\, \rm\AA} \le \lambda_r \le 1600 {\, \rm\AA}$ is $S/N > 0.5$.

We define the \lya forest region as the rest-frame wavelength range
$1041 {\, \rm\AA } \le \lambda_r \le 1185 {\, \rm\AA }$, 
and use the `value-added' co-added spectra made publicly available with DR9 (\cite{2013AJ....145...69L}), removing any 
pixels with the bit mask or the sky mask set, as defined in 
\cite{2013AJ....145...69L}. We also correct the pipeline estimate of the 
noise variance using the recipe described in \cite{2013AJ....145...69L}.

Lines of sight that include no
more than one detected DLA from the ``DLA concordance catalogue''
in \cite{Carithers2013} are included in the analysis,
after masking the central region of the DLA absorption line and
correcting for the damped wings outside this central region,
as explained in \cite{2013AJ....145...69L}.
Since the catalogue of \cite{Carithers2013} is not complete
(especially in low S/N spectra), some residual contamination of DLAs
is expected.



\section{Method}
\label{sec:method}

  The method we use to measure the quasar-\lya cross-correlation is the
same one as for the cross-correlation of Damped Lyman Alpha systems
(DLAs) with the \lya absorption, described in \cite{2012JCAP...11..059F}.
Here, a brief summary of the method is presented, discussing in
detail only the issues that are special for our analysis of the quasar
cross-correlation and any differences with respect to the method
used for DLAs.

  The observed flux in each pixel of the quasar spectra is
$ f_i = C_i \, \bar F(z_i) \left[1 + \delta_{Fi} \right] + N_i$, where
$C_i$ is the quasar continuum (equal to the flux that would be observed
in the absence of absorption), $\bar F(z_i)$ is the mean transmitted
fraction, $\delta_{Fi}$ is the \lya transmission fluctuation, and $N_i$
is the observational noise, which is assumed to be uncorrelated in all
pixel pairs. To estimate $\delta_{Fi}$ and its cross-correlation we
must first model the continua of the quasars. This is done using the
PCA technique described in \cite{2012AJ....143...51L}, but without
applying the {\it Mean Flux Regulation} described in this paper, which can
suppress the large-scale power in the \lya forest in ways that are
difficult to model. In the same way as in \cite{2012JCAP...11..059F}, we
generally apply instead the {\it Mean Transmission Correction}
(hereafter MTC), which enforces
the mean transmission in each quasar spectrum to be equal to the value
measured in independent observations by \cite{2008ApJ...681..831F}, 
using equations (3.5) and (3.6) in \cite{2012JCAP...11..059F}. This
correction is useful to remove the broadband noise caused by
spectrophotometric errors, but we will also show results when no
correction to the quasar continua is applied.
We will see in section \ref{sec:results} that including the MTC
increases the accuracy of the measured quasar bias in our parameterized
model by $\sim 30 \%$. However, this correction distorts the
cross-correlation function and the fitted theoretical model must 
be corrected to take this effect into account (see Appendix A in 
\cite{2012JCAP...11..059F}).

  The mean transmitted fraction $\bar F_i$ is measured here in 150 bins
of $\Delta z=0.01$ over the range $1.9 < z < 3.4$ (no \lya forest data
is used outside this range). These redshift bins are three times smaller
than the ones used in \cite{2012JCAP...11..059F} for the same purpose.
The measured $\bar F(z_i)$ has fluctuations when using small
redshift bins owing to systematic errors in the calibrating reference
stars (\cite{2013A&A...552A..96B}, \cite{2013AJ....145...69L}),
and we found that the fine redshift bins
are necessary to correctly eliminate the effect of these fluctuations.
In general, the larger number of available quasars compared to DLAs
allows for a more accurate measurement of the cross-correlation for the
quasars, and therefore greater care needs to be taken in the analysis
for quasars.

  The cross-correlation is computed with the simple estimator 
(see Appendix B of \cite{2012JCAP...11..059F}
for a discussion of the approximations involved and the differences
with an optimal estimator):
\begin{equation}
 \hat\xi_{A} = \frac{\sum_{i \in A}  w_i\, \delta_{Fi}}{\sum_{i \in A} w_i} ~,
 \label{eq:xiA}
\end{equation}
where the summation is done over all quasars and over all \lya pixels
that are within a bin ($A$) of the separation from each quasar in the 
perpendicular ($\sigma$) and parallel ($\pi$) directions, $\delta_{Fi}$ is 
the estimated value of the transmission fluctuation from the observed flux 
and the continuum model, and the weights $w_i$ are computed independently at
each pixel from the noise $N_i$ provided for the DR9 data
\citep{2012AJ....144..144B}, assuming a model for the intrinsic \lya 
absorption variance that is added in quadrature to the noise 
(equation 3.10 in \cite{2012JCAP...11..059F}).

  A set of 9 bins in $\sigma$ and 18 bins in $\pi$ are used to measure
the cross-correlation, which are the same ones as in
\cite{2012JCAP...11..059F} except that we add the
extra bin at large separations $60 \hmpc < \sigma < 80 \hmpc$, and
the same for both signs of $\pi$ (the cross-correlation is measured
without assuming symmetry with respect to a sign change of the parallel
separation $\pi$). This procedure yields a total of 162 bins.
The weighted average values of $(\pi,\sigma$) of all the contributing
pixel pairs to every bin $A$ are computed together with the
cross-correlation from equation (\ref{eq:xiA}),
using the same weights.  These averages are generally close but not
exactly equal to the central values of each bin. The models to be
fitted are evaluated at these weighted averages of $(\sigma,\pi)$.
A single redshift bin is generally used for the mean redshift $z$ of
the \lya forest pixel and the quasar, which is required to be in the range
$2.0 < z < 3.5$, although some results are also presented in the next
section using three redshift bins for the purpose of testing redshift
evolution.

  The errors of the cross-correlation are computed in two different
ways: 1) The covariance matrix is estimated assuming
Gaussianity and a model power spectrum for the \lya forest intrinsic
autocorrelation. 2) Our quasar sample is divided into
twelve subsamples in adjacent areas of the sky, and the
cross-correlation is computed separately in each subsample to infer
bootstrap errors. The method used and the subsample areas are the same
as in \cite{2012JCAP...11..059F}. The covariance matrix is computed
including only pixel pairs up to a transverse separation $\sigma < 5 \hmpc$
in order to make the computer time required for the calculation more
manageable. Most of the contributions to the covariance comes from
pixel pairs in the same line of sight ($\sigma=0$).

  We use the same linear theory model described in Section 3.6 of
\cite{2012JCAP...11..059F} (their equation 3.16), with bias parameters
$b_q$ and $b_F$, and redshift distortion parameters $\beta_q$ and
$\beta_F$, for the quasars and the \lya forest, respectively. 
The quasar redshift distortion parameter obeys the relation
$\beta_q=f(\Omega)/b_q$, and for the \lya forest we impose the condition
$b_F(1+\beta_F)=-0.336$ from the observational 
result of \cite{2011JCAP...09..001S} obtained from the measured \lya
autocorrelation at $z=2.25$.
We also impose that $\beta_F$ and $b_q$ are constant
with redshift and $b_F \propto (1+z)^{2.9}$, as discussed in 
\cite{2011JCAP...09..001S}.
However, whereas the effect of redshift errors for DLAs could be
neglected, this is not the case for quasars. We therefore
add two extra free parameters to the model: a dispersion $\epsilon_z$
and a mean offset $\Delta_z$ of the quasar redshift error, assuming
these errors to be Gaussian. The theoretical model for the linear
cross-correlation is smoothed with a Gaussian in the parallel direction
with this dispersion and offset.

  Fits to this model are generally done using only separation bins
at $r=(\sigma^2+\pi^2)^{1/2} > 15 \hmpc$ in order to avoid the near
region that is possibly
affected by radiation and non-linear effects, although some fits will
also be shown using all bins at $r > 7 \hmpc$. The theoretical
prediction is corrected for the MTC using the equations explained in
Appendix A of \cite{2012JCAP...11..059F}.

\section{Results}
\label{sec:results}

\subsection{Fiducial model}

\begin{figure}[h!]
 \begin{center}
  \begin{tabular}{ccc} 
   \includegraphics[scale=0.4, angle=-90]{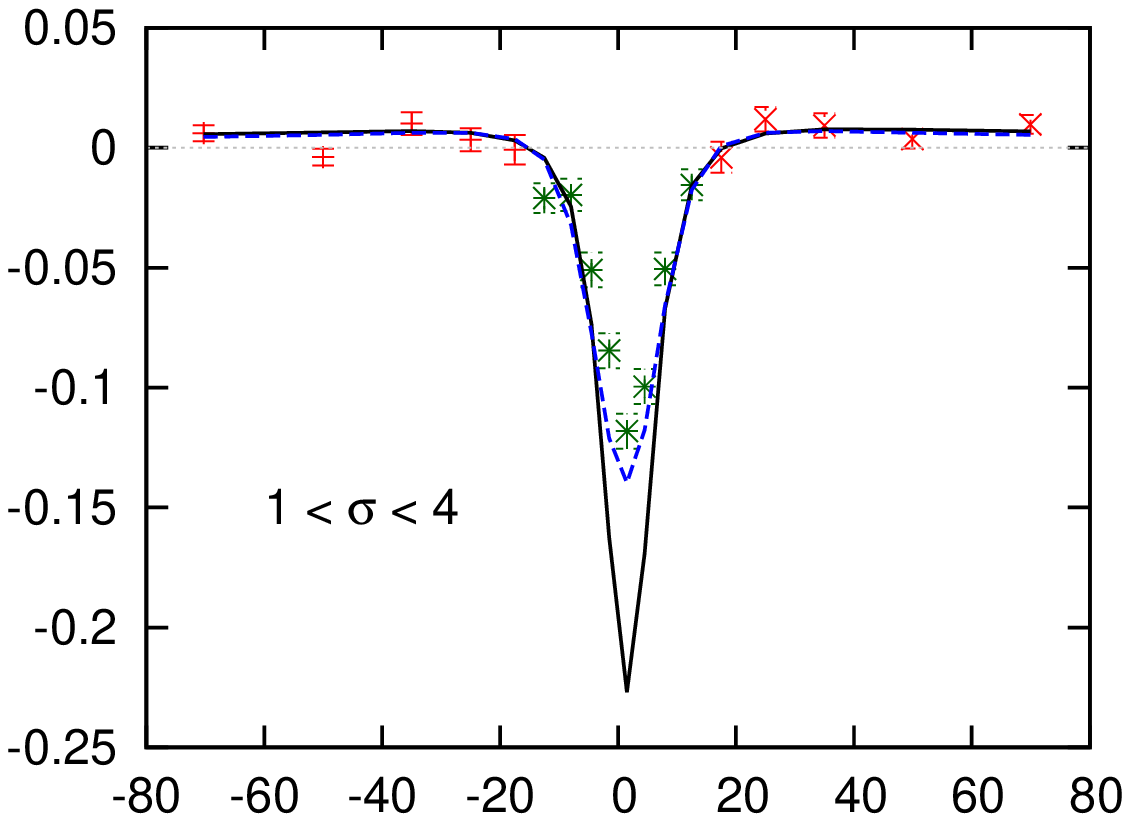} &
   \includegraphics[scale=0.4, angle=-90]{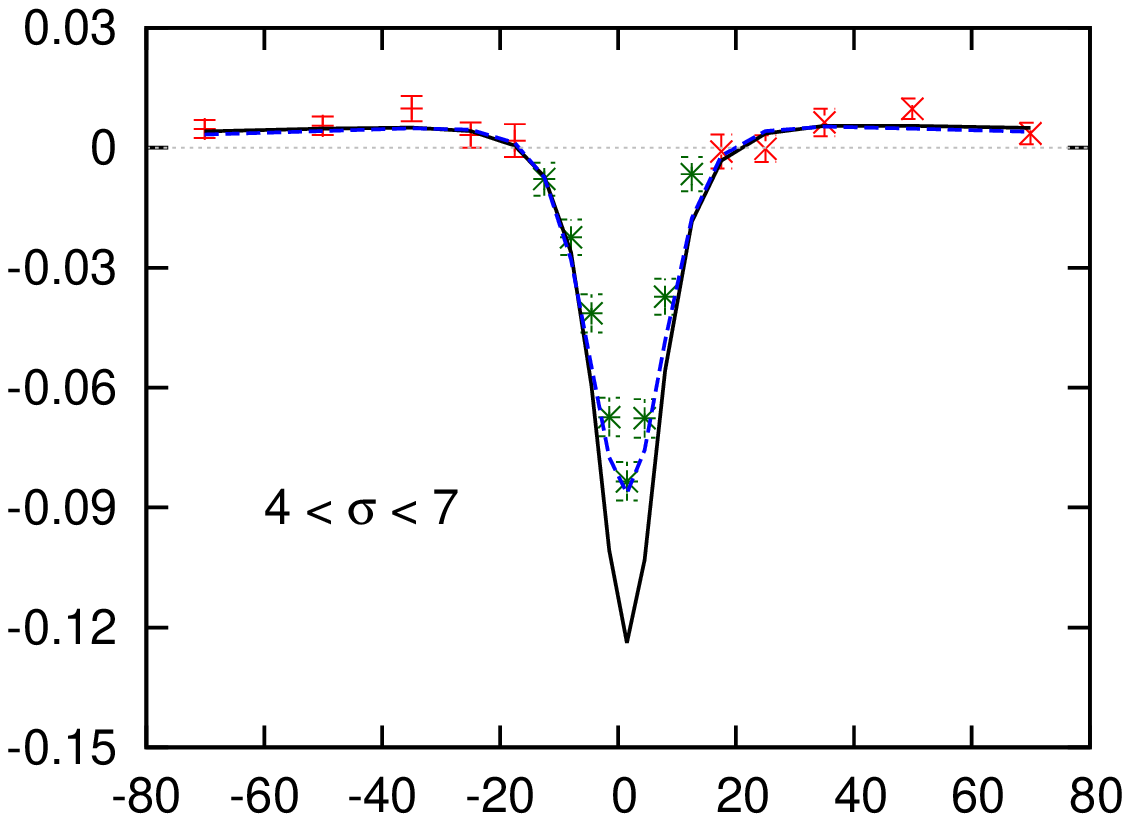} &
   \includegraphics[scale=0.4, angle=-90]{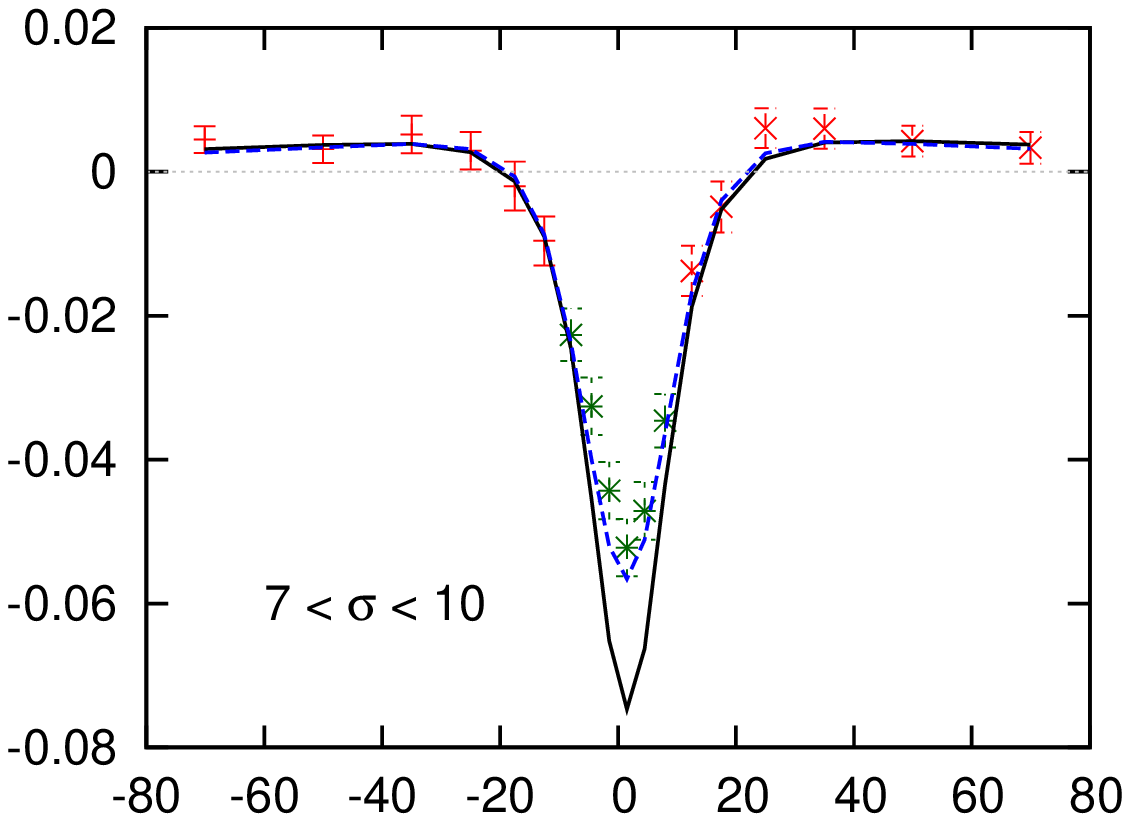} \\ 
   \includegraphics[scale=0.4, angle=-90]{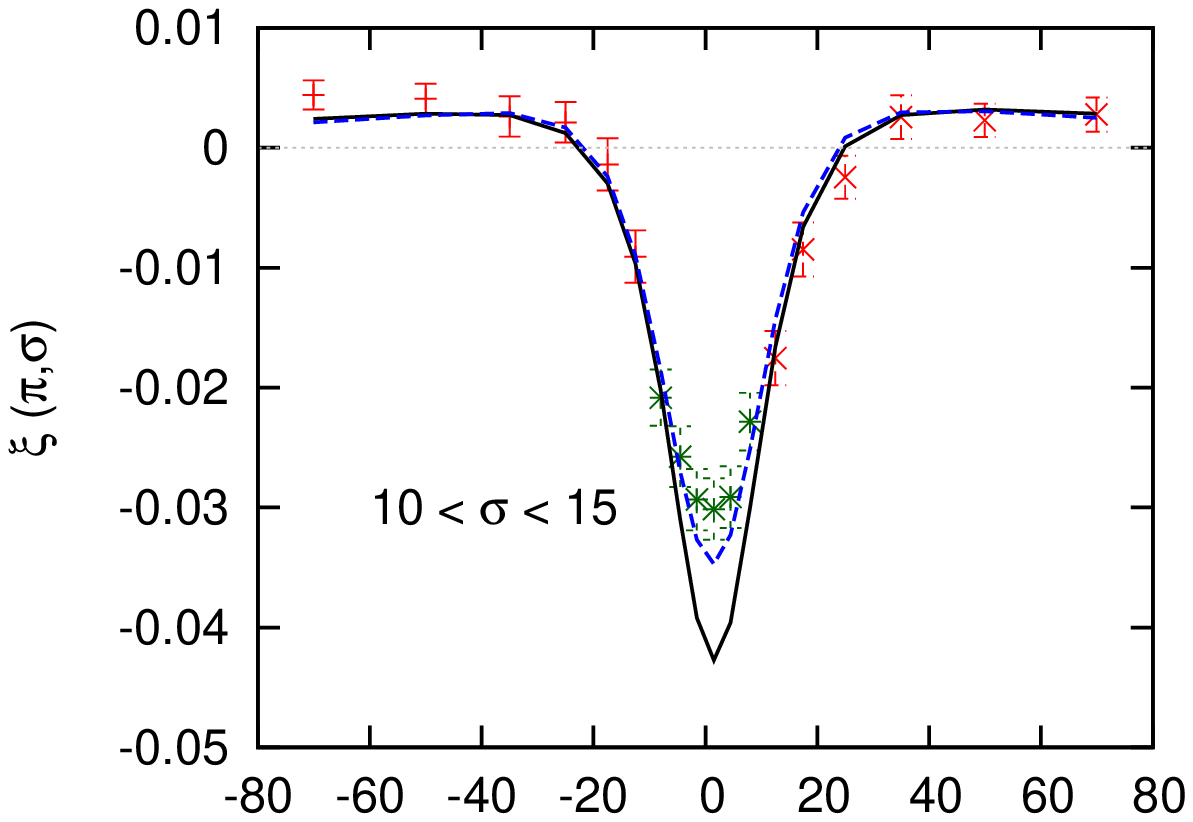} &
   \includegraphics[scale=0.4, angle=-90]{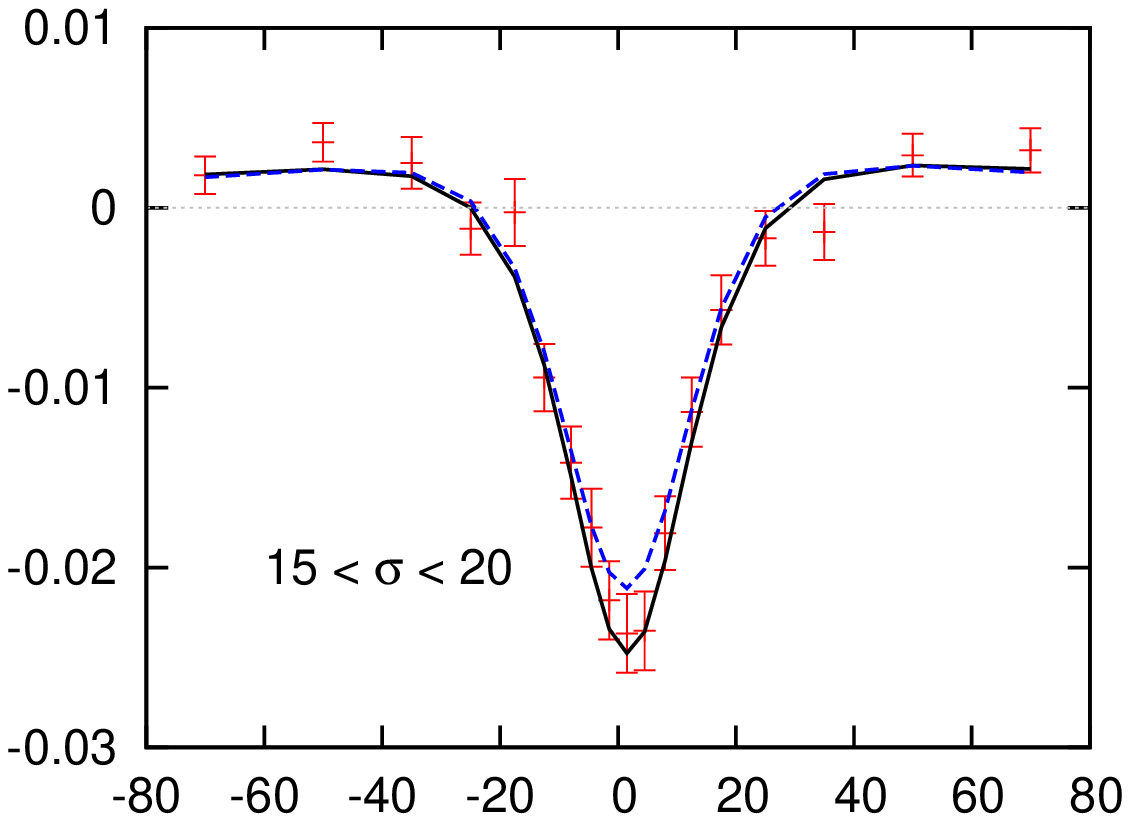} &
   \includegraphics[scale=0.4, angle=-90]{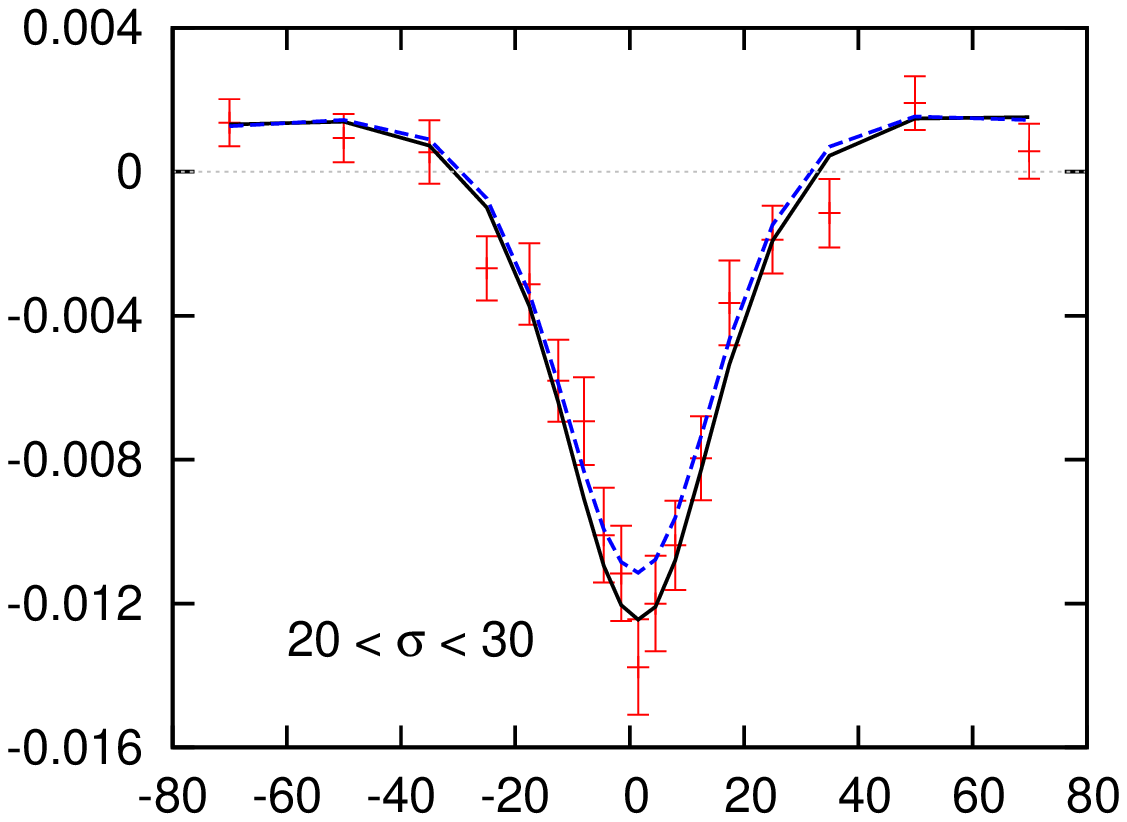} \\
   \includegraphics[scale=0.4, angle=-90]{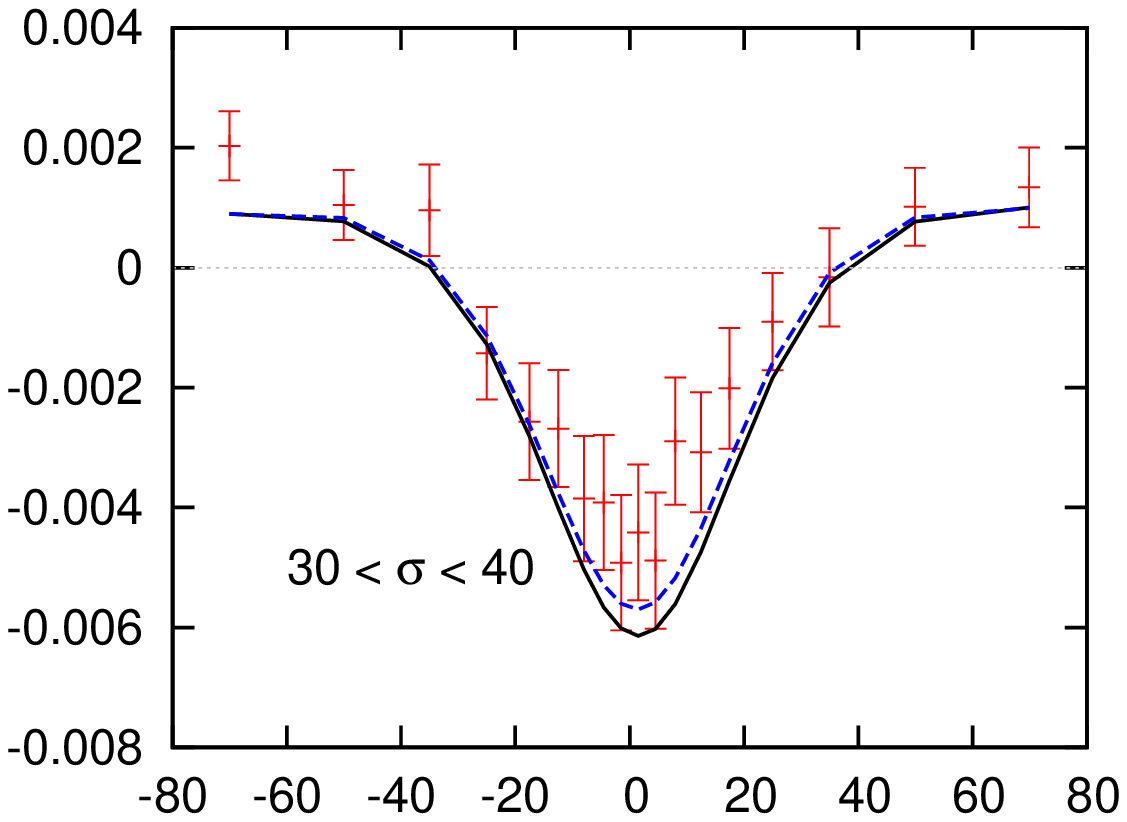} & 
   \includegraphics[scale=0.4, angle=-90]{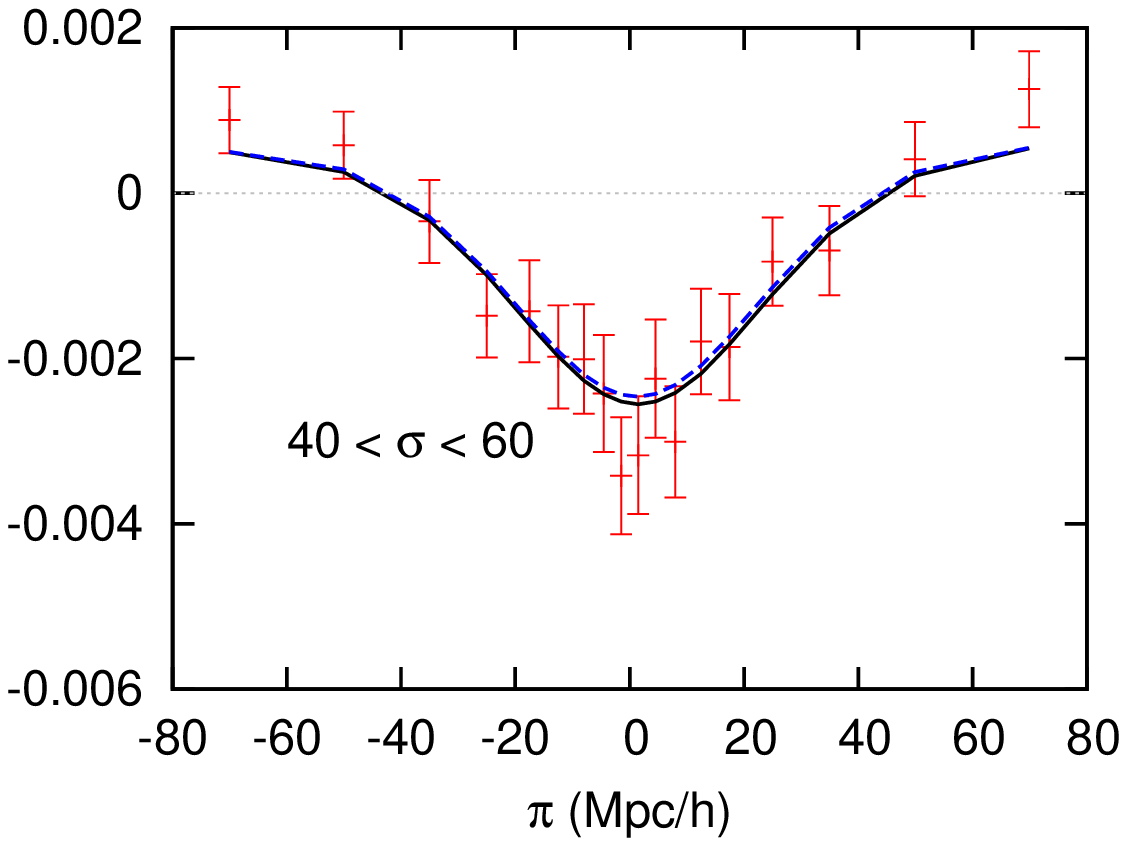} &
   \includegraphics[scale=0.4, angle=-90]{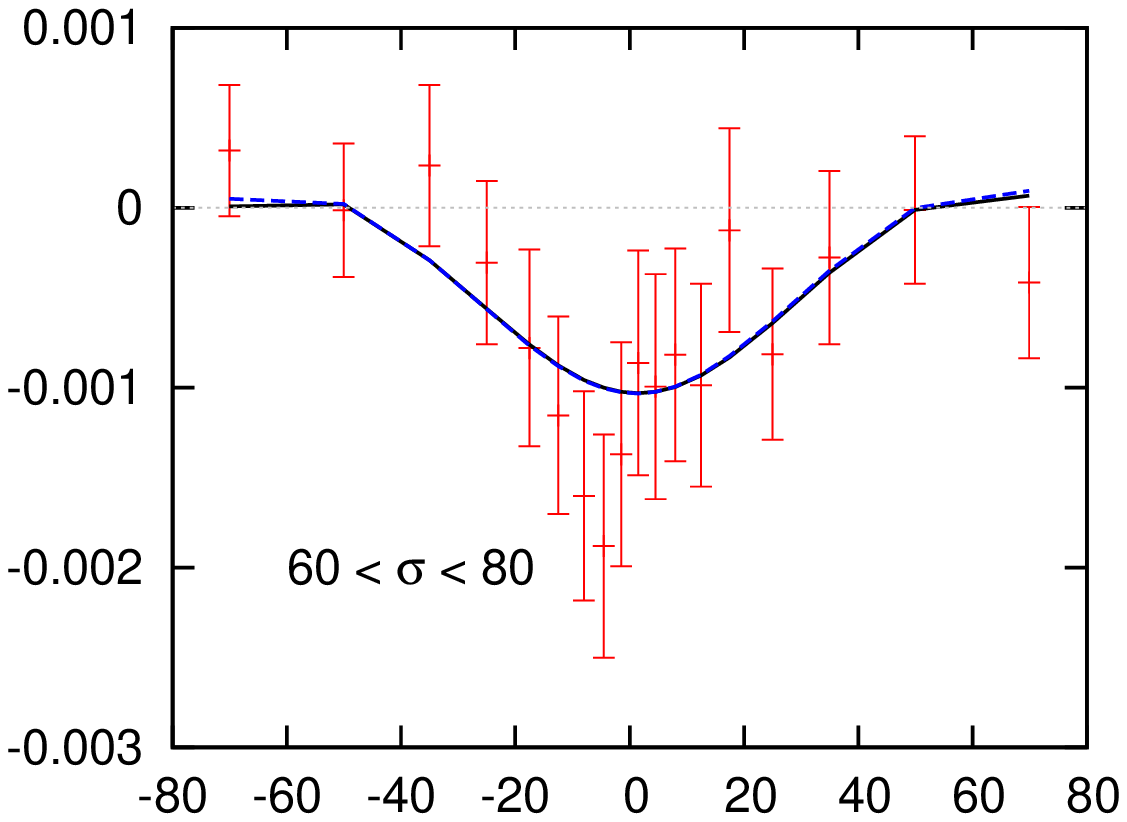} \\
  \end{tabular}
 \end{center}
 \caption{Measured cross-correlation in the indicated bins of
  perpendicular separation $\sigma$, as a function of the parallel
  separation $\pi$. The data points in green have a total separation 
  $r=(\sigma^2 + \pi^2)^{1/2}<15 \hmpc$, and are not used in most of our fits.
  Solid (dashed) dark (blue) lines show the best fit model for the fiducial 
  analysis, when using bins with separations down to $r=15 \hmpc$ ($r=7 \hmpc$).
 }
 \label{fig:cross}
\end{figure}


The results of the cross-correlation of the \lya transmission with
quasars, with the method presented in section \ref{sec:method}, are
shown in figure \ref{fig:cross} for each bin in the transverse
separation $\sigma$. The error bars are the diagonal elements of the
computed covariance matrix, which we have found to be consistent with
the bootstrap errors from the scatter of the cross-correlation function
measured in subsamples. Two model fits to the data are also plotted
in figure \ref{fig:cross}: the solid (black) line uses only bins with
separation $r=(\sigma^2+\pi^2)^{1/2} > 15 \hmpc$,
and the dashed (blue) line uses all the
bins with $r > 7 \hmpc$. The data points in green have a separation 
$r < 15 \hmpc$ and are not used in our main analysis.
The same results are shown as contour plots in the left panel of figure
\ref{fig:cont2D}, with our best fit fiducial model, the one using only
the $r > 15 \hmpc$ bins, shown in the right panel.

\begin{figure}[h!]
 \begin{center}
  \includegraphics[scale=0.37]{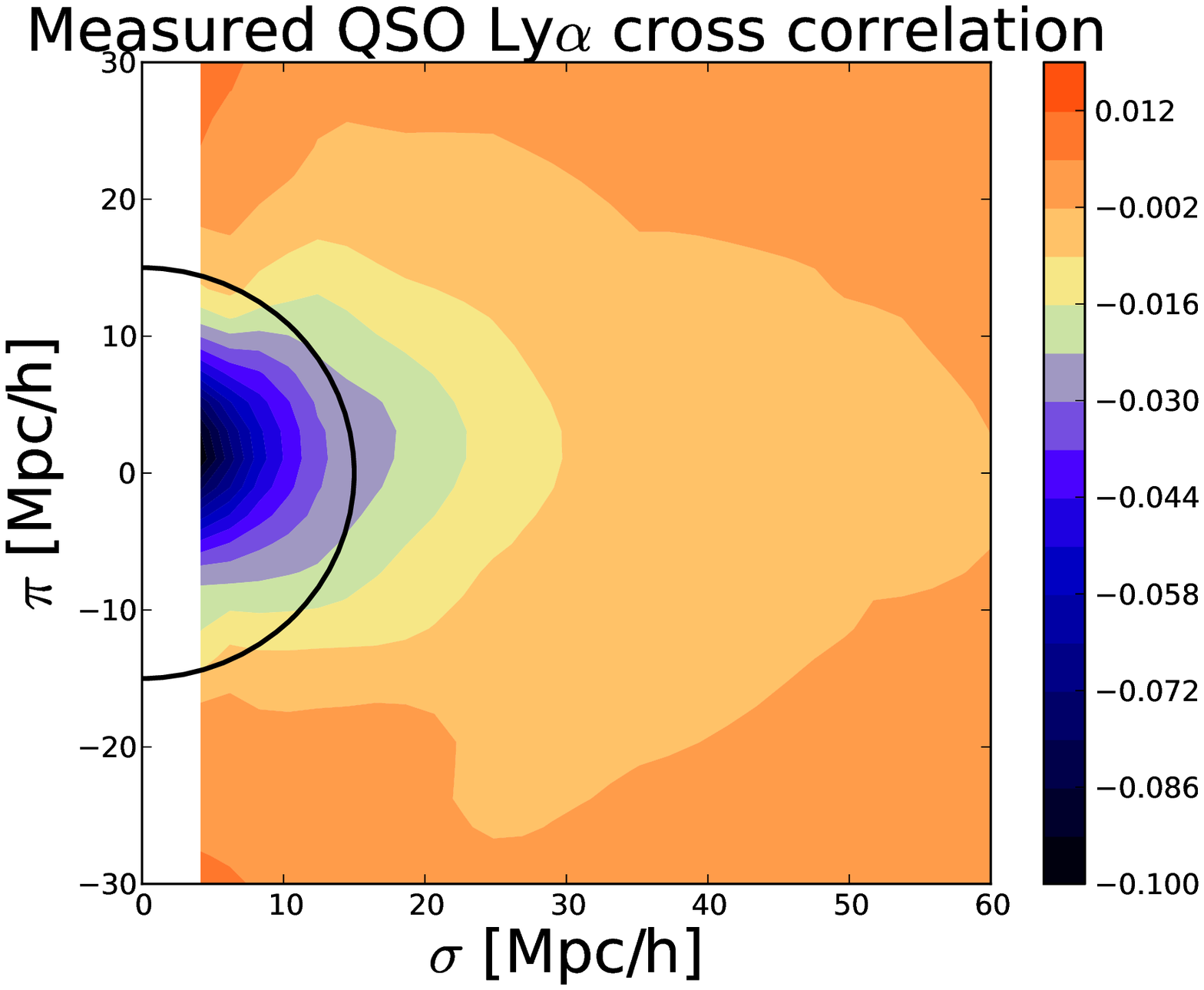}
  \includegraphics[scale=0.37]{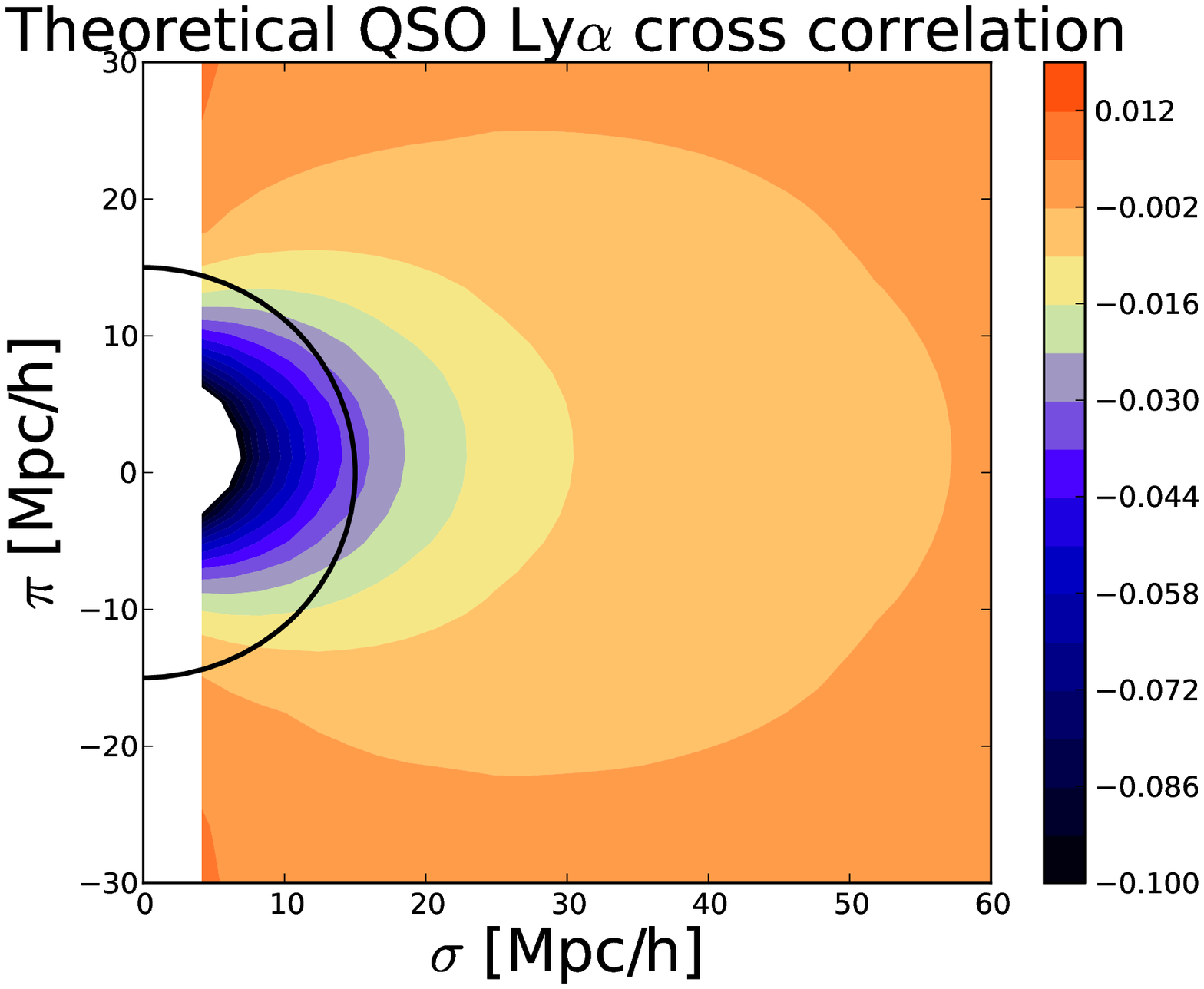} 
 \end{center}
 \caption{Two dimensional contours of the measured cross-correlation 
  (left panel), compared to the best fit theoretical models for 
  $r>15\hmpc$ (right panel). The black circle corresponds to $r=15 \hmpc$.}
 \label{fig:cont2D}
\end{figure}

  Our fiducial model fit applies the MTC to the data and includes the
corresponding correction to the theory, and uses the whole sample of
61342 quasars described in section \ref{sec:dataset} with the PCA quasar
redshifts. The
model has four free parameters, as described at the end of section
\ref{sec:method}: $\beta_F$ (the \lya redshift space distortion
parameter), $b_q$ (the quasar bias factor), $\epsilon_z$ (the rms of the
quasar redshift error distribution), and $\Delta_z$ (the mean offset of
the quasar redshift), while the \lya forest bias parameter $b_F$ is
computed for each value of $\beta_F$ using the well-constrained quantity 
$b_F(1+\beta_F)=-0.336$ at $z=2.25$ from the \lya autocorrelation result
of \cite{2011JCAP...09..001S}.
The first row in table \ref{tab:results} gives the parameters for the
best fit result (in the sense of minimum value of $\chi^2$) using 
the covariance matrix as described earlier, with the value of $\chi^2$
and the number of degrees of freedom given in the last column. Errors
correspond to the contours of $\Delta\chi^2=1$, after marginalizing
over all other parameters. In the case of
$\epsilon_z$, only an upper limit is provided when $\epsilon_z=0$ is
within the $\Delta\chi^2=1$ contour. The parameters for other models
that will be described below are given in the additional rows of the table.
The variables related to quasar redshift errors, $\epsilon_z$ and $\Delta_z$,
are expressed in units of $\kms$, reflecting the directly-measured
separation along the line of sight.






\begin{table}
 \centering
  \begin{tabular}{c|ccccc}
     & $\beta_F$  & $b_q$   & $\epsilon_z (\kms)$  & $\Delta_z (\kms)$  & $\chi^2$ (d.o.f) \\
\hline
 FIDUCIAL	& $1.1^{+0.17}_{-0.15}$ & $3.64^{+0.13}_{-0.15}$ & $<370$ & $-157^{+38}_{-36}$  & 116 (130) \\
 $r > 7 \hmpc$ & $1.67^{+0.23}_{-0.18}$ & $3.34^{+0.12}_{-0.16}$ & $433^{+44}_{-60}$ & $-136^{+20}_{-19}$  & 164 (152) \\
 NOCOR		& $3.38^{+0.68}_{-0.77}$ & $4^{+0.15}_{-0.18}$ & $591^{64}_{-120}$ & $-147^{+33}_{-32}$  & 142 (130) \\
 NOMTC		& $0.67^{+0.18}_{-0.11}$ & $3.58^{+0.18}_{-0.15}$ & $<280$ & $-152^{+34}_{-37}$  & 112 (130) \\
\hline
 LOW-Z & $1.81^{+0.56}_{-0.44}$ & $3.79^{+0.31}_{-0.33}$ & $660^{+137}_{-167}$ & $-134^{+70}_{-54}$  & 124 (130) \\
 MID-Z & $1.14^{+0.26}_{-0.26}$ & $3.34^{+0.21}_{-0.24}$ & $<450$ & $-115^{+63}_{-50}$  & 131 (130) \\
 HIGH-Z & $1.19^{+0.35}_{-0.25}$ & $3.88^{+0.29}_{-0.3}$ & $<428$ & $-226^{+70}_{-62}$  & 134 (130) \\
\hline
 LOW-L & $1.09^{+0.29}_{-0.18}$ & $3.65^{+0.25}_{-0.2}$ & $<340$ & $-151^{+58}_{-56}$  & 128 (130) \\
 MID-L & $1.03^{+0.36}_{-0.32}$ & $3.29^{+0.27}_{-0.24}$ & $<630$ & $-113^{+56}_{-61}$  & 113 (130) \\
 HIGH-L & $1.41^{+0.51}_{-0.24}$ & $4.21^{+0.26}_{-0.26}$ & $394^{+170}_{-220}$ & $-189^{+46}_{-69}$  & 118 (130) \\
\hline
 Z\_VISUAL	& $1.2^{+0.23}_{-0.16}$ & $3.65^{+0.14}_{-0.16}$ & $399^{+110}_{-99}$ & $-231^{+28}_{-38}$  & 142 (130) \\
 Z\_PIPELINE	& $1.13^{+0.21}_{-0.21}$ & $3.4^{+0.15}_{-0.16}$ & $546^{+86}_{-100}$ & $-154^{+43}_{-24}$  & 114 (130) \\
 Z\_CIV		& $1.34^{+0.19}_{-0.17}$ & $3.66^{+0.13}_{-0.15}$ & $503^{+72}_{-79}$ & $-412^{+28}_{-36}$  & 137 (130) \\
 Z\_CIII	& $1.44^{+0.26}_{-0.21}$ & $3.5^{+0.2}_{-0.17}$ & $648^{+87}_{-67}$ & $-436^{+48}_{-34}$  & 137 (130) \\
 Z\_MgII 	& $1.73^{+0.44}_{-0.39}$ & $3.55^{+0.26}_{-0.19}$ & $636^{+110}_{-150}$ & $-79^{+38}_{-56}$  & 126 (130) \\
  \end{tabular}
  \caption{Best fit parameters and $\chi^2$ for the different analyses:
    FIDUCIAL (with the MTC and the corrected theory, using $r>15\hmpc$),
    $r > 7 \hmpc$ (extending to smaller scales),
    NOCOR (MTC, uncorrected theory),
    NOMTC (PCA-only continuum fitting, uncorrected theory),
    data split in redshift bins (LOW-Z for $2 < z < 2.25$, MID-Z
    for $2.25 < z < 2.5$, HIGH-Z for $2.5 < z < 3.5$), 
    data split in quasar absolute magnitude (LOW-L for
    $-25.2 < M_i < -23$, MID-L for $-26.1 < M_i < -25.2$, and
    HIGH-L for $-30 < M_i < -26.1$)
    and finally different quasar redshift estimates
    (Z\_VISUAL,Z\_PIPELINE,Z\_CIV, Z\_CIII,Z\_MgII).
    Uncertainties correspond to values with $\Delta \chi^2 =1$, with upper
    limits for $\epsilon_z$ when $\Delta\chi^2 < 1$ at $\epsilon_z=0$.}
  \label{tab:results}
\end{table}

  Our basic result, seen in the figures and table \ref{tab:results},
 is that the simple linear theory model for the
cross-correlation of quasars as tracers of the mass distribution and
the \lya forest provides an excellent fit to all the data at large
scales. Moreover, the predicted redshift distortions are an excellent
match to the observed cross-correlation, as seen in figure
\ref{fig:cont2D}, confirming the large-scale mass inflow toward the
quasar host halos expected from the gravitational evolution of density
perturbations. The quasar bias factor required to match the
cross-correlation is $b_q=3.64^{+0.13}_{-0.15}$, in excellent agreement with
the independently-determined bias factor from the quasar
auto-correlation, $b_q=3.8 \pm 0.3$, from \cite{2012MNRAS.424..933W}.
The redshift distortion parameter of the \lya forest is found to be
$\beta_F=1.1^{+0.17}_{-0.15}$, also in good agreement with the measurement of
\cite{2011JCAP...09..001S}. Finally, we find that the best match of the
quasar redshift error distribution requires a significant mean
offset of $\Delta_z = -160 \kms$, with the negative sign indicating
that the PCA quasar redshifts are on average too small (so the
cross-correlation seen in figure \ref{fig:cont2D} is shifted to a
redshift higher than that of the quasar), and a surprisingly small
error dispersion of $\epsilon_z < (370, 490) \kms$ at the
$\chi^2= (1,4)$ confidence levels. Note that $\epsilon_z$ is the
combination of the observational error of the quasar redshift and
the intrinsic velocity dispersion of quasars with respect to their
host halo. We need to keep in mind, however, that this upper limit
is obtained using only the pixels at $r>15 \hmpc$, and that the
value of $\epsilon_z$ has degeneracies with slight modifications
of our simple four-parameter fiducial model.



\begin{figure}[h!]
 \begin{center}
  \begin{tabular}{cc}
   \includegraphics[scale=0.4]{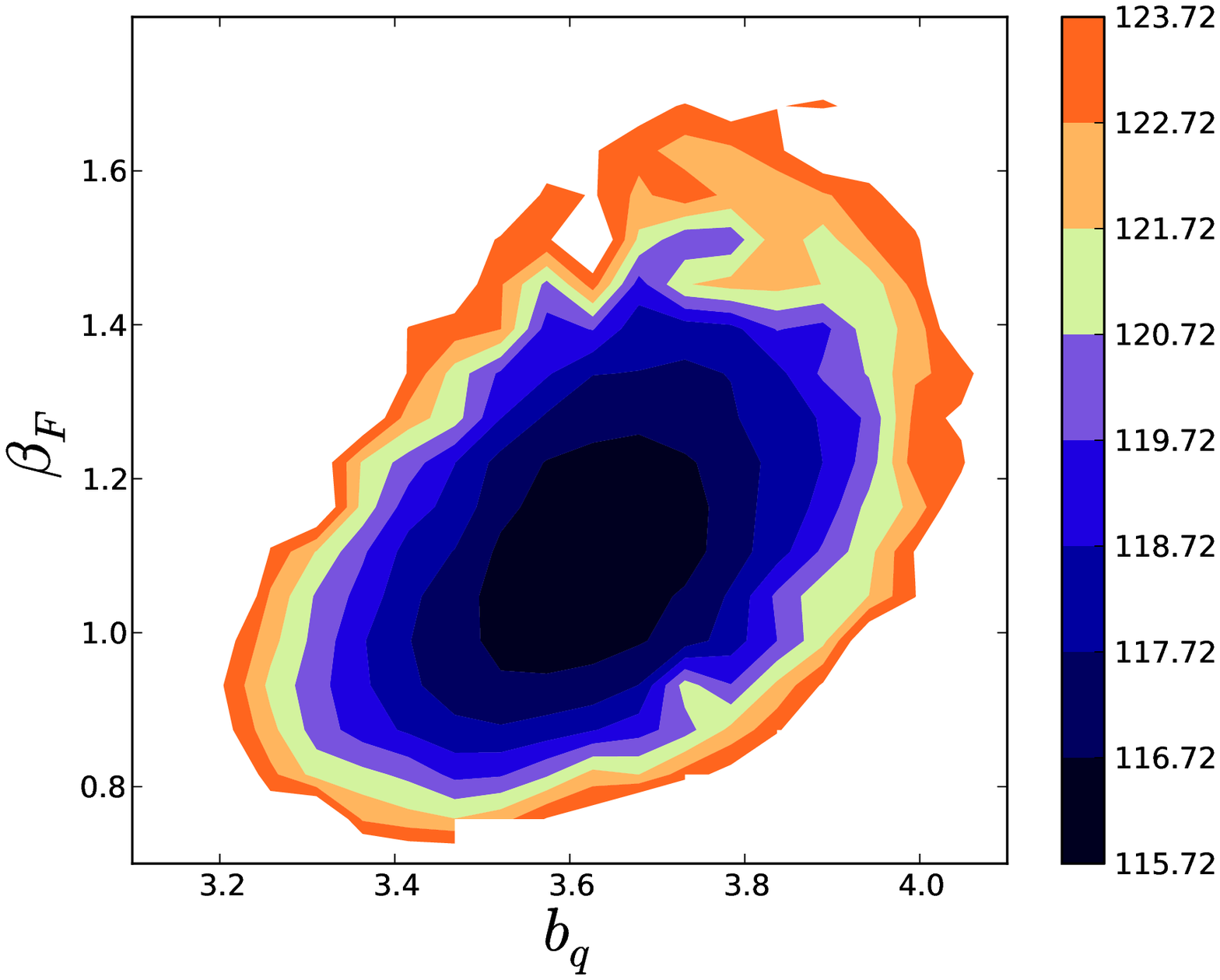} &
   \includegraphics[scale=0.4]{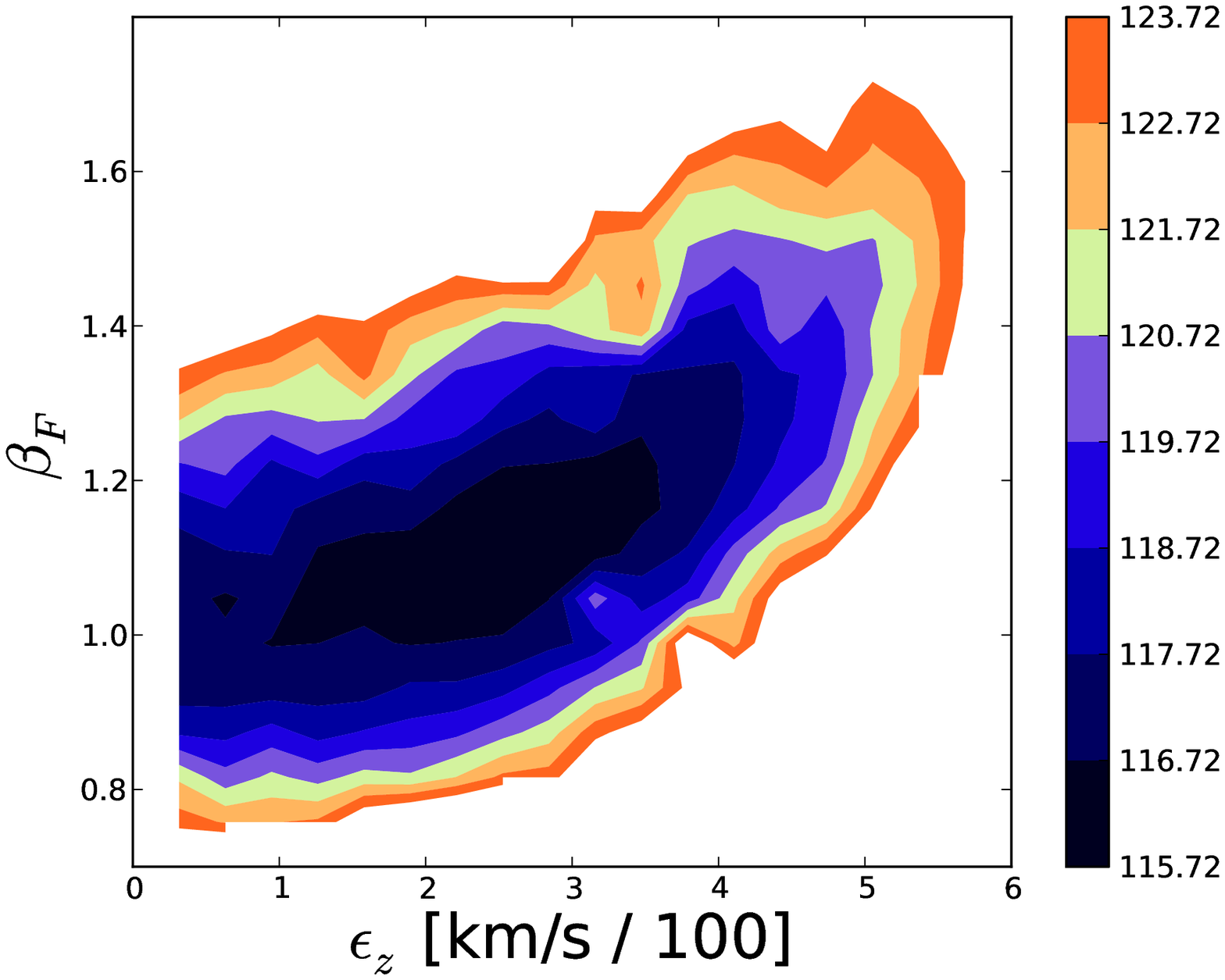} 
  \end{tabular}
 \end{center}
 \caption{Contours of $\chi^2$ in the two-parameter plane of the
  \lya forest redshift distortion parameter versus:
  quasar bias (left), and
  quasar redshift error dispersion (right). The number of degrees of freedom
  is 130.}
 \label{fig:degen}
\end{figure}


  This degeneracy is well illustrated in our model by the $\chi^2$
contours for the parameters $\beta_F$ and $\epsilon_z$, shown in figure
\ref{fig:degen} (right panel). The quadrupole moment of the cross-correlation is
determined mostly by these two parameters. The quadrupole moment
increases with $\beta_F$ and decreases with $\epsilon_z$, but the effect
of $\epsilon_z$ is obviously important only at small radius (the
dependence of the quadrupole moment on the quasar bias factor is 
relatively small
because of the requirement $\beta_q=f(\Omega)/b_q$, which implies a
small value for $\beta_q$). A modification of our model at small radius
owing to the radiation effects of the quasars or non-linearities may
modify the best fit values of $\epsilon_z$ and $\Delta_z$.
The left panel of figure \ref{fig:degen} also demonstrates that the quasar
bias $b_q$ tends to slightly increase with increasing $\beta_F$ for a
given analysis.

  The value of $\chi^2=116$ for our fiducial model fit for 130 degrees
of freedom, and the agreement of the fitted parameters with other
independent determinations, shows that the observed cross-correlation
is sufficiently well reproduced without the need to include the effect
of the quasar ionizing radiation. Clearly, the linear
overdensity around the quasar host halo is the dominant effect when
measuring the quasar-\lya cross-correlation away from the line of
sight, on large scales and for the luminosities of the BOSS quasars.

  The second row of table \ref{tab:results} gives the best fit parameters
when all the bins at separations down to $r > 7 \hmpc$ are used. The
best fit shifts to a larger redshift distortion parameter of the \lya
forest, a lower quasar bias factor, and a larger quasar redshift error
dispersion. The $\chi^2$ worsens significantly, with an increase
of 48 when adding only 22 degrees of freedom. This result suggests that our
4-parameter model does not include all the important physical effects
when analyzing the entire range of separations in figure \ref{fig:cross}.
Non-linearities and radiation effects are likely to play a role at small
separations, and a more complex analysis 
will be required to discern this.



\subsection{Bootstrap errors}

In general, we have tested that the bootstrap errors computed as
described in section \ref{sec:method} are in agreement with the errors 
derived from the covariance matrix.
We mention here as an example this error comparison
for the quasar bias factor, when keeping the other parameters 
fixed to their best fit value of our fiducial analysis.
The covariance matrix yields errors from the
$\Delta \chi^2=1$ contour of $b_q=3.634^{+0.138}_{-0.138}$, 
and the result of 100 bootstrap realizations from the 12 subsamples of 
our data set (see \cite{2012JCAP...11..059F}) is $b_q = 3.657 \pm 0.135$,
in perfect agreement with the standard analysis. In the rest of the paper 
we use the uncertainties set by the contours of $\Delta \chi^2 =1$.

\subsection{Scale dependence of the quasar bias factor}

\begin{figure}[h!]
  \begin{center}
   \includegraphics[scale=0.70, angle=-90]{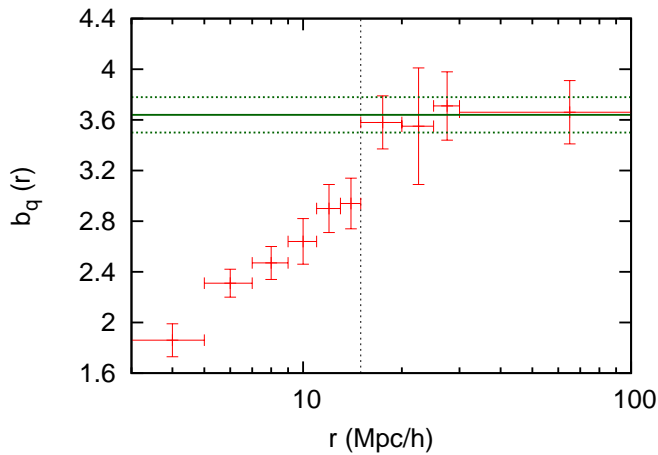}
  \end{center}
  \caption{Fitted QSO bias $b_q$ in several bins of the separation $r$,
   when fixing the other parameters to their best fit value in the
   fiducial model, $\beta_F=1.1$,
   $\epsilon_z=2.42 \hmpc$, $\Delta_z=-1.57 \hmpc$. The horizontal lines
   show the best fit
   and uncertainties when using all separations larger than $r>15\hmpc$, 
   the scale that is marked with a vertical dotted line in the figure.}
 \label{fig:bias_r}
\end{figure}

In figure \ref{fig:bias_r} we present the value of the quasar bias that is
obtained when using only bins within narrow rings of the separation $r$.
For this analysis, we fix the other parameters to their best fit value
when using the range $15 \hmpc < r < 100 \hmpc$, which are
$\beta_F=1.1$, $\epsilon_z = 2.42 \hmpc$, $\Delta_z=-1.57 \hmpc$.
The horizontal lines in the figure show
the best fit and uncertainties when using all separations above $r>15 \hmpc$, 
i.e., all the points lying to the right of the dotted vertical line.

  The constancy of the bias factor for $r> 15 \hmpc$ is again a
success of the simple linear theory model for large scales, meaning
that the radial dependence of the cross-correlation agrees with the
prediction of the standard $\Lambda$CDM model of structure formation.
The smaller value of $b_q$ at smaller separation confirms our previous
conclusion that other effects are likely to be important at
$r < 15 \hmpc$.

  The amplitude of the cross-correlation is proportional to 
$b_F b_q \sigma_8^2$, this is the actual quantity we are
measuring, and the value for $b_q$ plotted in figure \ref{fig:bias_r}
assumes that $b_F(1+\beta_F)=-0.336$ at $z=2.25$ (corrected to the
mean quasar redshift $\bar z_q=2.38$, where $b_F(1+\beta_F)=-0.376$),
and $\sigma_8=0.8$.

\subsection{Impact of the Mean Transmission Correction (MTC)}

  Our fiducial model uses the {\it Mean Transmission Correction} (MTC)
as part of the continuum fitting, and corrects the theoretical model
accordingly by using the analytical expression derived in appendix A of 
\cite{2012JCAP...11..059F}. As a test of the importance of this
correction, table \ref{tab:results} gives the results for two additional
models: the NOCOR case (fourth row) treats the observations in the same
way as the fiducial model (applying the MTC to the data), but does not
correct the theoretical model. The result is a considerably worse fit
($\Delta \chi^2 = 26$), confirming the validity and the need for
the theoretical correction.
In the NOMTC case, the data for the transmission fluctuation is obtained
with the direct use of the PCA continua, without applying the MTC.
The uncorrected theory then fits the data properly, but the
errors of the best fit parameters are considerably worse, supporting
our reasons to use the MTC. 
An even smaller upper limit for $\epsilon_z$ is obtained for the
NOMTC case, indicating that this upper limit is questionable because of
its degeneracy with other model variations.

\begin{figure}[h!]
  \begin{center}
   \includegraphics[scale=0.60, angle=-90]{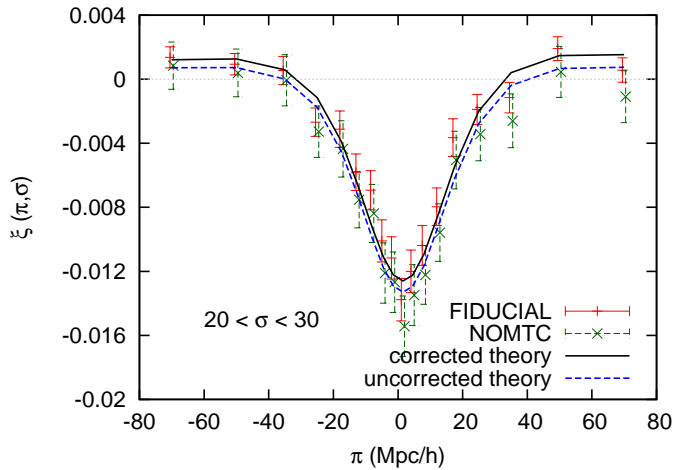}
  \end{center}
  \caption{Measured cross-correlation in the sixth bin in $\sigma$ 
   ($20 \hmpc < \sigma < 30 \hmpc$). Red (green) data points correspond to the 
   analysis with (without) the MTC step in the continuum fitting. The black 
   solid (blue dashed) line shows the best fit model for the fiducial 
   analysis with (without) the MTC correction.}
 \label{fig:MTC}
\end{figure}

Figure \ref{fig:MTC} shows the effect of the MTC on the measured 
cross-correlation, for the sixth bin of $\sigma$ 
($20 \hmpc < \sigma < 30 \hmpc$), as well as the correction we apply to the 
theoretical model. As seen in the plot, the theoretical correction clearly 
captures the difference in the analyses. The errorbars in the NOMTC analysis 
are considerably larger, due to the large spectro-photometric errors present in 
BOSS quasars. Since these errors have a coherent effect on all pixels of a 
given spectrum, the errorbars in the NOMTC analysis are strongly correlated.

\subsection{Evolution with redshift and variation with quasar luminosity}

We test here for the dependence of our measured quasar bias on the redshift
and quasar luminosity. In table \ref{tab:results} we show the results
obtained when splitting the quasar sample in three redshift bins: 
LOW-Z (2.00 $<$ z $<$ 2.25), MID-Z (2.25 $<$ z $<$ 2.50) and 
HIGH-Z (2.50 $<$ z $<$ 3.50). There is no clear evidence of evolution in the 
quasar bias parameter $b_q$, although our errors are large and the
redshift range that is probed is limited.

In the same table we show the results obtained when splitting the sample in
three luminosity bins: LOW-L for $-25.2 < M_i < -23.0$,
MID-L for $-26.1 < M_i < -25.2$, and HIGH-L for $-30.0 < M_i < -26.1$. 
Again, the changes are not significant, and unfortunately the quasar
redshift error distribution may vary with the quasar luminosity,
making it difficult to search for
any physical dependence of the cross-correlation with quasar luminosity
because of the parameter degeneracies.

\subsection{Variation with the quasar redshift estimator}
\label{ss:redshifts}

There are six different quasar redshift estimators specified in the DR9 quasar 
catalog (\cite{2012A&A...548A..66P}). In the main part of this study we use
the PCA redshift estimator (Z\_PCA in DR9Q), but we also show in table 
\ref{tab:results} the results obtained when using the other ones:
the visual inspection redshift (Z\_VI), the estimator from the BOSS pipeline 
(Z\_PIPELINE), and three estimators that use a single metal absorption 
line (Z\_CII,Z\_CIV,Z\_MgII). 

Table \ref{tab:results} shows the best fit parameters for each of the 
quasar redshift estimators, and the quasar bias obtained is consistent among
them. However, the redshift error distribution varies considerably for the
different estimators. Our fits suggest that all the estimators systematically
underestimate the quasar redshift by several hundreds of $\kms$,
except for the one based on the MgII line which is basically consistent
with $\Delta_z=0$, but has a larger dispersion $\epsilon_z$.

 The presence of a mean redshift displacement can also be tested using narrow 
emission lines of the quasar host galaxies, namely OII and OIII. The mean 
redshifts of the MgII and OII lines were compared by \cite{2010MNRAS.405.2302H}
in a large sample of low redshift SDSS quasars, who found that the difference 
had to be smaller than $30 \kms$. This comparison, however, would be difficult 
to carry out for the higher redshift quasars of our sample, where the OII 
line is shifted to the infrared.

  We caution, however, that these values of $\Delta_z$ and $\epsilon_z$
are likely to change once our simple four-parameter model is improved
with more parameters to include the quasar ionizing radiation effects.
The systematic negative value of $\Delta_z$ might be adjusting a real
asymmetry of the cross-correlation introduced by the reduction of the
effective quasar luminosity with the time delay. The effect of the
time-delayed quasar ionizing radiation on the cross-correlation should
obviously have a different radial dependence than a simple
shift of the quasar redshifts, but some degree of degeneracy can be
expected. At the same time, it seems difficult to believe that
the true dispersion $\epsilon_z$, arising from both observational errors
and the intrinsic velocity dispersion of quasars within halos, is
smaller than $\sim 500 \kms$, and its value will also probably be
modified in a more complex model that better reflects the underlying
physical effects.

\section{Effects of the Quasar Radiation}
\label{sec:discussion}

  As we have seen in the previous section, the large-scale form of the
quasar-\lya cross-correlation is consistent with the linear overdensity
expected around the quasar host halos. This result may be surprising because
of the previous detection of the proximity effect when measuring this
cross-correlation along the line of sight, caused by the increased
ionization of the intergalactic medium induced by the quasar ionizing
radiation, although the BOSS quasars are of lower luminosity than most
of the quasars on which the proximity effect was measured and
therefore the expected additional ionization is much weaker in our
case compared to previous studies. In this section we present a simple
estimate of the expected radiation effect on the full three-dimensional
cross-correlation with our quasar sample. A more detailed analysis
involving fitting of a more complex model that includes both the
radiation and overdensity effects is left for a future study.

  The radiation of a quasar of luminosity $L_\nu$ increases the
photoionization rate of gas at a proper distance $d$ by the amount
\begin{equation}
 \Gamma_q={1\over 4\pi d^2 h} \int_{\nu_{HI}}^\infty d\nu\, {L_\nu
  \sigma_{HI}(\nu)\over \nu} ~,
\end{equation}
where $\nu_{HI}$ is the hydrogen Lyman limit frequency, $\sigma_{HI}$ is
the photoionization cross section, and $h$ is the Planck constant. We
neglect here the absorption by Lyman limit systems and the redshifting
of the radiation, which reduces the quasar intensity when $d$ is
comparable to the absorption mean free path or the local horizon.
Let the relative fluctuation of the photoionization rate
relative to its average value $\Gamma_0$ be $\delta_\Gamma$. In general,
$\delta_\Gamma$ can be affected by many sources. Neglecting the effects
of ionizing source clustering (which increases the mean ionizing flux
near a quasar beyond that emitted by the quasar itself), assuming
a quasar spectrum $L_\nu \propto \nu^{-\alpha}$, and approximating
$\sigma_{HI}(\nu) \propto \nu^{-3}$ at $\nu > \nu_{HI}$,
the average value of $\delta_\Gamma$ near the quasar is
\begin{equation}
\label{eq:phr}
  \delta_\Gamma = {\Gamma_q\over \Gamma_0} \simeq
 {L_{\nu_{HI}} \sigma_{HI}(\nu_{HI}) \over
 4\pi d^2 h \Gamma_0 (3+\alpha)} ~.
\end{equation}

  The impact of the perturbation $\delta_\Gamma$ on the \lya forest
can be calculated using the approximation that the gas is purely
photoionized (neglecting any contribution from collisional ionization)
and that the neutral fraction is everywhere much smaller than unity.
In this case, the optical depth at any point in the spectrum of the
\lya forest is inversely proportional to the photoionization rate, so
the fractional perturbation in the optical depth is $\delta_\tau =
1/(1+\delta_\Gamma)-1 \simeq - \delta_\Gamma$, where the last approximate
equality assumes $\delta_\Gamma \ll 1$. Now, let $F_0=e^{-\tau_0}$ be the
\lya transmission fraction when the photoionization rate has the uniform
value $\Gamma_0$, with a distribution $P(F_0)$. The transmission in the
presence of radiation fluctuations is
$F=e^{-\tau_0(1-\delta_\Gamma)}$, and the mean transmission is
\begin{align}
 \bar F &= \int_0^1 dF_0\, P(F_0)\, e^{-\tau_0(1-\delta_\Gamma)} \\
 & \simeq \int_0^1 dF_0\, P(F_0)\, F_0 (1+\tau_0\delta_\Gamma) \nonumber \\
 & \simeq \bar F_0 - \delta_\Gamma \int_0^1 dF_0\, P(F_0)\, 							F_0\, \log(F_0) ~. \nonumber 
\end{align}
We can now define the radiation bias factor of the \lya forest,
$b_\Gamma$, as the linear variation of the mean transmission
fluctuation $\delta_F$ in response to a fractional variation
$\delta_\Gamma$ in the radiation intensity, analogously to the density
and peculiar velocity gradient bias factors. Hence, near a quasar the
mean transmission fluctuation will vary by $\delta_F = b_\Gamma
\delta_\Gamma$ owing to the fractional radiation perturbation
$\delta_\Gamma$, where 
\begin{equation}
 b_\Gamma = - \frac{1}{\bar F} \int_0^1 dF_0\, P(F_0)\, F_0\, \ln(F_0) ~.
\end{equation}

If we use as the transmission distribution a log-normal function in the
optical depth, with mean $\bar F_0 =0.8$ and dispersion
$\sigma_F=0.124$, which is close to the observed distribution at
$z=2.3$, we find $b_\Gamma=0.13$.

  We now estimate the average intensity produced by the quasars in our
sample. The mean flux per unit frequency of a quasar of magnitude $g$ is
$f_\nu=10^{-0.4(48.6+g)}$, expressed in cgs units (\cite{1996AJ....111.1748F}).
The mean value of $f_\nu$ we obtain for our quasar sample
is $f_\nu=2.4\times 10^{-28}\erg\seg^{-1}\cm^{-2}\hz^{-1}$ at the
center of the $g$-band (at $4800\,{\rm \AA}$; this corresponds to a $g$
magnitude of 20.5)
At the mean redshift
$\bar z_q=2.38$ of our quasar sample, the implied mean luminosity at
the shifted $g$-band center $\lambda=1420\, {\rm \AA}$ is $L_{\nu} =
 4\pi D_L^2 f_\nu/(1+\bar z_q)=3.1\times 10^{30}\erg\seg^{-1}\hz^{-1}$.
We assume a mean spectral slope from this wavelength to the Lyman
limit wavelength $L_\nu\propto \nu^{-1}$ 
(\cite{2001AJ....122..549V}, \cite{2012ApJ...752..162S}) , which results in
$L_{\nu_{HI}}=2.0\times 10^{30} \erg\seg^{-1}\hz^{-1}$.
The corresponding quasar flux at a characteristic comoving distance of
interest of $d(1+\bar z_q) = 20 \hmpc$, or proper distance $d=8.33 \mpc$, 
is then $f_{\nu_{HI}} = 2.4\times 10^{-22} \erg\seg^{-1}\cm^{-2}\hz^{-1}$.
Finally, using equation (\ref{eq:phr}) with the spectral index $\alpha=1.5$ for
$\nu > \nu_{HI}$, the derived
photoionization rate is $\Gamma_q=5.1\times 10^{-14}\seg^{-1}$.

  If we assume a mean photoionization rate from the average of all
sources $\Gamma_0=10^{-12}\seg^{-1}$, then at this comoving distance
of $20\hmpc$, we have $\delta_\Gamma\simeq 0.05$, and the mean
perturbation on the transmission should be $\delta_F\simeq b_\Gamma
\delta_\Gamma \simeq 0.0065$. In the absence of any absorption effects,
this perturbation should vary in proportion to $d^{-2}$.

  Comparing to figure \ref{fig:cont2D}, we see that this radiation effect 
should displace the value of the cross-correlation by one contour at
$r=(\sigma^2+\pi^2)^{1/2}=20 \hmpc$. In the third panel of figure 
\ref{fig:cross} (for the range $7 \hmpc < \sigma < 10 \hmpc$), the increase 
of the cross-correlation would be $\sim 0.03$ at small $\pi$, which is larger
than the difference between the data points and our fiducial model fit,
and in the fifth panel (for the range $15 \hmpc < \sigma < 20 \hmpc$),
the increase would be $\sim 0.008$. These changes would significantly affect
our fiducial fit, implying a substantial increase of $b_q$ in order to 
compensate for the radiation effect to a value $b_q\simeq 5$, which would 
disagree with the autocorrelation measurement of \cite{2012MNRAS.424..933W}. 
We therefore suspect that the quasar radiation effect is reduced relative 
to this simple estimate.

  The effect of the ionizing radiation is not expected to be
isotropic around the quasar in redshift space. Quasars are likely to
emit their light anisotropically, with lower intensity near the plane of
the accretion disk around the black hole. A flux limited sample of
quasars preferentially selects sources with their axis (the direction
of brightest emission) near our line of sight, so on average the flux
from quasars in the perpendicular direction should be lower than in the
parallel direction. Time variability of the quasars also implies that
the mean quasar ionizing flux affecting the \lya forest depends on the
time delay, $ct_d=r+\pi$: the \lya forest is illuminated by the
luminosity of the quasar at a time $t_d$ before the epoch when we are
observing. The selection effect again causes the average quasar in a
flux-limited sample to have
lower luminosity in the past compared to the present, and this should
introduce an asymmetry depending on the sign of $\pi$. 
A hint of this
signature of the radiation effect is seen in figure \ref{fig:cross}, in the 
region $\pi\simeq -20 \hmpc$ and $\sigma < |\pi|$, which is consistent with the
increase of $\delta_F\simeq 0.007$ we have estimated at $r=20 \hmpc$
from the radiation effect. If this hint is correct, quasars might shine
at close to
the expected luminosities within $\sim 45 ^\circ$ of the line of sight
and for time delays $t_d < 10^7$ years, and have effective lower
luminosities at larger angles from the line of sight and for longer time
delays.

  In any case, obtaining solid conclusions on the contribution of the
radiation effect to the cross-correlation and the statistical
significance of any detection requires a more detailed modeling with more
fitting parameters, additional data and a more careful inclusion of all the
important effects. We plan to perform this study in the future when the
BOSS survey is completed. We mention here, however, that the radiation
effects we have discussed may be altering the value of the linear model
parameters we have fitted. In particular, the redshift offset
$\Delta_z$ may in part be the result of the attempt to adjust the
radiation and time-delay effect which introduces the asymmetry depending
on the sign of $\pi$.

 Non-linear effects in the clustering of both quasars and \lya absorption, 
as well as the non-linearity in the relation between optical depth and 
transmitted flux fraction, may also have an impact on the cross-correlation 
at small scales. However, we compared the linear theory predictions to the 
results of numerical simulations of \cite{2008MNRAS.387..377K} for the 
cross-correlation, and we found that non-linearities become important only at 
scales smaller than a few $\hmpc$, whereas the radiation effects are a more 
important correction at intermediate scales of $\sim 10 \hmpc$.

  Finally, we note that HeII reionization may alter the IGM temperature near 
quasars because of the additional heating involved (\cite{1994MNRAS.266..343M},
\cite{2009ApJ...694..842M}), which may result in an additional effect on the 
cross-correlation. The effect would likely also vary as the inverse squared 
distance from the quasar. However, the duration of the HeII reionization is 
probably much longer than the typical lifetime of a quasar, implying that most 
of the temperature fluctuations originated from the HeII reionization would 
be caused by quasars that have long been dead.

\subsection{Predictions for the proximity effect}

  Our fitted linear theory model for the mean overdensity around a
quasar can be used to make a prediction on the impact of this
overdensity on any measurements of the proximity effect of quasars on
the line of sight due to their ionizing radiation. In figure \ref{fig:prox}, 
we show the prediction for the line of sight cross-correlation, as
$\pi\xi(\sigma=0, \pi)$, for our two first models in table 1: 1) the
fiducial model fit using pixels at $r>15 \hmpc$ (solid, black line), and
2) the same model using all pixels at $r>7 \hmpc$ (dashed, blue line; 
in this subsection we choose the convention that $\pi$ is positive
even though for the three-dimensional discussion $\pi$ is negative in
front of the quasar).
Also shown as the red, dotted line is the expected radiation effect
in our sample of quasars, which we have plotted according to our simple
estimate above as $\xi=0.0065(20 \hmpc/\pi)^2$. Note that this curve is 
proportional to the mean quasar luminosity, so the proximity effect due to the 
radiation dominates over the overdensity effect for quasars of much greater 
luminosity than the ones in the BOSS sample.

\begin{figure}[h!]
  \begin{center}
   \includegraphics[scale=0.60, angle=-90]{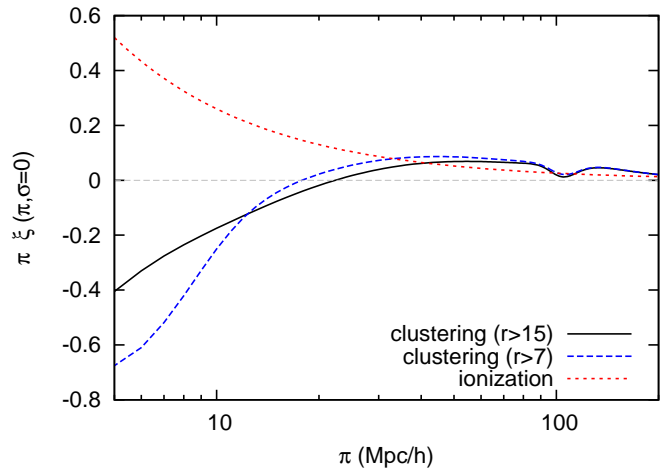}
  \end{center}
  \caption{Cross-correlation function along the line of sight, i.e., using 
   pixels and their background quasar. The black solid (blue dashed) line 
   shows our best fit theory for $r>15 \hmpc$ ($r>7 \hmpc$). The dotted red 
   line shows the expected radiation effect.
  }
 \label{fig:prox}
\end{figure}

  As shown in this figure, the overdensity effect should roughly
cancel the radiation effect at $\pi\sim 10 \hmpc$ for the
luminosities of typical BOSS quasars. Most studies of the proximity
effect have used brighter quasars to obtain higher signal-to-noise
spectra, but figure \ref{fig:prox} provides a correction that should be applied
to any future measurements of the proximity effect due to radiation
under the assumption that the quasar bias factor does not depend
on quasar luminosity. For $\pi > 20 \hmpc$, the overdensity effect
causes an increased $\delta_F$ (decreased absorption) that is therefore
added to any radiation effect, because the peculiar velocity
gradient due to the infall of matter toward the quasar halo is a more
important effect than the overdensity.

\section{Conclusions}
\label{sec:conclusions}

We have measured the cross-correlation of quasars and the \lya forest
transmission in redshift space using more than 60000 quasars from BOSS.
This unprecedentedly large number of quasar spectra for such a study has 
allowed a statistically significant detection of this cross-correlation out to
separations of $\sim 80 \hmpc$, and a detailed determination of its
radial and angular dependence. The cross-correlation is consistent with
the linear theory prediction of the standard $\, \rm\Lambda$CDM model, with
the expected redshift distortions, on scales $r > 15\hmpc$. Fitting
these large-scale measurements to a linear model with four parameters,
we find that the BOSS quasars at $\bar z_q=2.38$ have a mean bias
factor $b_q = 3.64^{+0.13}_{-0.15}$. This result is consistent with the
quasar bias measured from the auto-correlation of a sub-sample of the
same quasars, $b_q = 3.8 \pm 0.3$, as presented in \cite{2012MNRAS.424..933W}.
The halo mass having this bias factor is $M_h\simeq 3\times 10^{12} \hMs$. 
This measurement can be compared to the halo mass corresponding to
the mean bias factor inferred for DLAs, the other population of objects
for which the bias factor was measured from the same approach measuring
the cross-correlation with the \lya forest in BOSS, which was $M_h\simeq
4\times 10^{11} \hMs$, although a large scatter in halo masses is
expected for both populations of objects and only the mean bias factor
is measured. We do not detect any dependence of quasar bias on luminosity
or redshift, also consistent with the results from \cite{2012MNRAS.424..933W}.

  The large-scale fit to the cross-correlation also provides a
measurement of the \lya forest redshift distortion parameter,
$\beta_F = 1.1^{+0.17}_{-0.15}$. This value is $\sim 1.5 \sigma$ higher than 
the one measured in \cite{2011JCAP...09..001S}, reducing the tension
with previous numerical simulations (\cite{2003ApJ...585...34M}). 
The value of $\beta_F$ measured from this cross-correlation has the
advantage of being less sensitive to systematic errors in the
spectrophotometric calibration, although it should be equally affected
by the presence of DLAs, Lyman limit systems and metal lines in the
spectra which tend to decrease $\beta_F$ (\cite{2012JCAP...07..028F}).

  The cross-correlation at scales $r< 15\hmpc$ is not well fitted by
the simple linear theory model we have used, because its amplitude is
lower than expected in the model fitted to large scales, as seen in figure 
\ref{fig:bias_r}. We have argued that a likely explanation is the effects of
the ionizing radiation of the quasars, which are of the right order of
magnitude to explain this discrepancy for the luminosity of the BOSS
quasars. If all the BOSS quasars emitted their light isotropically
and with constant luminosities, the enhanced ionization of the
surrounding medium would also affect the
cross-correlation we have fitted at $r> 15 \hmpc$, implying a higher
quasar bias factor to compensate for this ionization effect. However, the
quasar radiation effect is likely to be decreased owing to anisotropic
emission and finite quasar lifetimes.

  The impact of the quasar ionization can also be studied by means of
the line of sight proximity effect, and using the cross-correlation
measured from especially targeted quasar pairs at smaller angular
separations than in the BOSS sample. 
In the future, these studies will need to model the superposed effects of 
the mean overdensity around the quasar host halos and the ionization effects.
Here, we have presented predictions from our fitted models for the
correction that needs to be applied to the line of sight proximity
effect for the mean overdensity around quasars, before attempting to
infer anything about the ionization effect of the quasar. This
correction has been neglected in the past (\cite{1982MNRAS.198...91C},
\cite{1986ApJ...309...19M},\cite{1988ApJ...327..570B},
\cite{2000ApJS..130...67S}), and is in fact small for the
most luminous quasars that have been used to study the proximity effect,
but becomes important for quasars of luminosities typical of the
BOSS sample. 

The expected redshift distortions also imply that the
effect of the overdensity is smaller on the line of sight compared to
the perpendicular direction, and changes sign at separations 
$|\pi| \gtrsim 20\hmpc$ owing to the induced peculiar velocity gradient.
We note that this predicted correction assumes linear theory, so it can be 
altered at small scales.

  The quasar-\lya cross-correlation can also be useful to constrain
the redshift errors of the quasars. Our results for the mean redshift
offset $\Delta z$ indicate that the quasar estimators that have been
used are systematically too low, except for the one based on MgII which
is closest to zero. We have warned, however, that the result for this
redshift offset may be affected by the quasar radiation effect with
finite quasar lifetimes, which can introduce an asymmetry depending on
the sign of $\pi$.

  Future studies of the quasar-\lya cross-correlation have a promising
potential for probing both the large-scale distribution of matter around
quasars and the characteristics of the ionizing emission. The
Baryon Acoustic Oscillation peak, detected recently in the \lya
autocorrelation (\cite{2013A&A...552A..96B},\cite{2013JCAP...04..026S}), can
also be detected in the quasar-\lya cross-correlation.
The ratio of the BAO peak amplitudes in the monopole,
$\xi^{BAO}_{q\alpha}/ \xi^{BAO}_{\alpha\alpha}$, should be equal to 
(\cite{HAMIL92})
\begin{equation}
{\xi^{BAO}_{q\alpha}\over \xi^{BAO}_{\alpha\alpha} } = {b_q \over b_F} \,
{ 1 + (\beta_F+\beta_q)/3 + \beta_F\beta_q/5 \over
  1 + 2\beta_F/3 + \beta_F^2/5 } ~.
\end{equation}
Using the values $b_q=3.7$, $\beta_F=1.1$, and $b_F=0.16$, we find
$\xi^{BAO}_{q\alpha}/ \xi^{BAO}_{\alpha\alpha}\simeq 18$, and using 
$\xi^{BAO}_{\alpha\alpha} \simeq 2\times 10^{-5}$ (\cite{2013A&A...552A..96B}), 
we expect $\xi^{BAO}_{q\alpha}\simeq 4\times 10^{-4}$. This BAO peak should be
detectable in the cross-correlation before the end of the BOSS survey 
(see figure \ref{fig:cross}). At the same
time, detailed modeling of the cross-correlation on a broad range of
scales will hopefully allow for the separation of the overdensity and
radiation effects, allowing for a measurement of the characteristic
anisotropy of quasar emission and of the lifetimes of quasars.

\begin{acknowledgments}

We would like to thank Rupert Croft, Joe Hennawi, Patrick McDonald, 
Matt McQuinn, Uro{\v s} Seljak, An\v{z}e Slosar and Nao Suzuki for very useful 
conversations and comments on preliminary versions of this publication.

This research used resources of the National Energy Research Scientific 
Computing Center (NERSC), which is supported by the Office of Science of 
the U.S. Department of Energy under Contract No. DE-AC02-05CH11231.
JM is supported in part by Spanish grant AYA 2009-09745.
 
Funding for SDSS-III has been provided by the Alfred P. Sloan
Foundation, the Participating Institutions, the National Science
Foundation, and the U.S. Department of Energy Office of Science.
The SDSS-III web site is http://www.sdss3.org/.

SDSS-III is managed by the Astrophysical Research Consortium for the
Participating Institutions of the SDSS-III Collaboration including the
University of Arizona,
the Brazilian Participation Group,
Brookhaven National Laboratory,
University of Cambridge,
Carnegie Mellon University,
University of Florida,
the French Participation Group,
the German Participation Group,
Harvard University,
the Instituto de Astrofisica de Canarias,
the Michigan State/Notre Dame/JINA Participation Group,
Johns Hopkins University,
Lawrence Berkeley National Laboratory,
Max Planck Institute for Astrophysics,
Max Planck Institute for Extraterrestrial Physics,
New Mexico State University,
New York University,
Ohio State University,
Pennsylvania State University,
University of Portsmouth,
Princeton University,
the Spanish Participation Group,
University of Tokyo,
University of Utah,
Vanderbilt University,
University of Virginia,
University of Washington,
and Yale University.

\end{acknowledgments}

\bibliography{cosmo,cosmo_preprints}

\begin{thebibliography}{10}
\providecommand*{\bibinfo}[2]{#2}
\providecommand*{\eprint}[1]{#1}
\providecommand*{\url}[1]{#1}
\bibitem{1971ApJ...164L..73L}
\bibinfo{author}{R.~{Lynds}}, \bibinfo{journal}{\apjl}
  \bibinfo{volume}{\textbf{164}}, \bibinfo{pages}{L73+} (\bibinfo{date}{Mar.
  1971}).
\bibitem{1998ARA&A..36..267R}
\bibinfo{author}{M.~{Rauch}}, \bibinfo{journal}{\araa}
  \bibinfo{volume}{\textbf{36}}, \bibinfo{pages}{267} (\bibinfo{date}{1998}),
  \eprint{arXiv:astro-ph/9806286}.
\bibitem{2006ApJS..163...80M}
\bibinfo{author}{P.~{McDonald}}, \bibinfo{author}{U.~{Seljak}},
  \bibinfo{author}{S.~{Burles}}, \bibinfo{author}{D.~J. {Schlegel}},
  \bibinfo{author}{D.~H. {Weinberg}}, \bibinfo{author}{R.~{Cen}},
  \bibinfo{author}{D.~{Shih}}, \bibinfo{author}{J.~{Schaye}},
  \bibinfo{author}{D.~P. {Schneider}}, \bibinfo{author}{N.~A. {Bahcall}},
  \emph{et~al.}, \bibinfo{journal}{\apjs} \bibinfo{volume}{\textbf{163}},
  \bibinfo{pages}{80} (\bibinfo{date}{Mar. 2006}),
  \eprint{arXiv:astro-ph/0405013}.
\bibitem{2011JCAP...09..001S}
\bibinfo{author}{A.~{Slosar}}, \bibinfo{author}{A.~{Font-Ribera}},
  \bibinfo{author}{M.~M. {Pieri}}, \bibinfo{author}{J.~{Rich}},
  \bibinfo{author}{J.-M. {Le Goff}}, \bibinfo{author}{{\'E}.~{Aubourg}},
  \bibinfo{author}{J.~{Brinkmann}}, \bibinfo{author}{N.~{Busca}},
  \bibinfo{author}{B.~{Carithers}}, \bibinfo{author}{R.~{Charlassier}},
  \emph{et~al.}, \bibinfo{journal}{\jcap} \bibinfo{volume}{\textbf{9}},
  \bibinfo{pages}{1} (\bibinfo{date}{Sep. 2011}), \eprint{1104.5244}.
\bibitem{PMN04}
\bibinfo{author}{C.~{Porciani}}, \bibinfo{author}{M.~{Magliocchetti}}, and
  \bibinfo{author}{P.~{Norberg}}, \bibinfo{journal}{\mnras}
  \bibinfo{volume}{\textbf{355}}, \bibinfo{pages}{1010} (\bibinfo{date}{Dec.
  2004}), \eprint{astro-ph/0406036}.
\bibitem{2005MNRAS.356..415C}
\bibinfo{author}{S.~M. {Croom}}, \bibinfo{author}{B.~J. {Boyle}},
  \bibinfo{author}{T.~{Shanks}}, \bibinfo{author}{R.~J. {Smith}},
  \bibinfo{author}{L.~{Miller}}, \bibinfo{author}{P.~J. {Outram}},
  \bibinfo{author}{N.~S. {Loaring}}, \bibinfo{author}{F.~{Hoyle}}, and
  \bibinfo{author}{J.~{da {\^A}ngela}}, \bibinfo{journal}{\mnras}
  \bibinfo{volume}{\textbf{356}}, \bibinfo{pages}{415} (\bibinfo{date}{Jan.
  2005}), \eprint{arXiv:astro-ph/0409314}.
\bibitem{2005MNRAS.360.1040D}
\bibinfo{author}{J.~{da {\^A}ngela}}, \bibinfo{author}{P.~J. {Outram}},
  \bibinfo{author}{T.~{Shanks}}, \bibinfo{author}{B.~J. {Boyle}},
  \bibinfo{author}{S.~M. {Croom}}, \bibinfo{author}{N.~S. {Loaring}},
  \bibinfo{author}{L.~{Miller}}, and \bibinfo{author}{R.~J. {Smith}},
  \bibinfo{journal}{\mnras} \bibinfo{volume}{\textbf{360}},
  \bibinfo{pages}{1040} (\bibinfo{date}{Jul. 2005}),
  \eprint{arXiv:astro-ph/0504438}.
\bibitem{2006ApJ...638..622M}
\bibinfo{author}{A.~D. {Myers}}, \bibinfo{author}{R.~J. {Brunner}},
  \bibinfo{author}{G.~T. {Richards}}, \bibinfo{author}{R.~C. {Nichol}},
  \bibinfo{author}{D.~P. {Schneider}}, \bibinfo{author}{D.~E. {Vanden Berk}},
  \bibinfo{author}{R.~{Scranton}}, \bibinfo{author}{A.~G. {Gray}}, and
  \bibinfo{author}{J.~{Brinkmann}}, \bibinfo{journal}{\apj}
  \bibinfo{volume}{\textbf{638}}, \bibinfo{pages}{622} (\bibinfo{date}{Feb.
  2006}), \eprint{arXiv:astro-ph/0510371}.
\bibitem{2007AJ....133.2222S}
\bibinfo{author}{Y.~{Shen}}, \bibinfo{author}{M.~A. {Strauss}},
  \bibinfo{author}{M.~{Oguri}}, \bibinfo{author}{J.~F. {Hennawi}},
  \bibinfo{author}{X.~{Fan}}, \bibinfo{author}{G.~T. {Richards}},
  \bibinfo{author}{P.~B. {Hall}}, \bibinfo{author}{J.~E. {Gunn}},
  \bibinfo{author}{D.~P. {Schneider}}, \bibinfo{author}{A.~S. {Szalay}},
  \emph{et~al.}, \bibinfo{journal}{\aj} \bibinfo{volume}{\textbf{133}},
  \bibinfo{pages}{2222} (\bibinfo{date}{May 2007}),
  \eprint{arXiv:astro-ph/0702214}.
\bibitem{2009ApJ...697.1634R}
\bibinfo{author}{N.~P. {Ross}}, \bibinfo{author}{Y.~{Shen}},
  \bibinfo{author}{M.~A. {Strauss}}, \bibinfo{author}{D.~E. {Vanden Berk}},
  \bibinfo{author}{A.~J. {Connolly}}, \bibinfo{author}{G.~T. {Richards}},
  \bibinfo{author}{D.~P. {Schneider}}, \bibinfo{author}{D.~H. {Weinberg}},
  \bibinfo{author}{P.~B. {Hall}}, \bibinfo{author}{N.~A. {Bahcall}},
  \emph{et~al.}, \bibinfo{journal}{\apj} \bibinfo{volume}{\textbf{697}},
  \bibinfo{pages}{1634} (\bibinfo{date}{Jun. 2009}), \eprint{0903.3230}.
\bibitem{2009ApJ...697.1656S}
\bibinfo{author}{Y.~{Shen}}, \bibinfo{author}{M.~A. {Strauss}},
  \bibinfo{author}{N.~P. {Ross}}, \bibinfo{author}{P.~B. {Hall}},
  \bibinfo{author}{Y.-T. {Lin}}, \bibinfo{author}{G.~T. {Richards}},
  \bibinfo{author}{D.~P. {Schneider}}, \bibinfo{author}{D.~H. {Weinberg}},
  \bibinfo{author}{A.~J. {Connolly}}, \bibinfo{author}{X.~{Fan}},
  \emph{et~al.}, \bibinfo{journal}{\apj} \bibinfo{volume}{\textbf{697}},
  \bibinfo{pages}{1656} (\bibinfo{date}{Jun. 2009}), \eprint{0810.4144}.
\bibitem{2012MNRAS.424..933W}
\bibinfo{author}{M.~{White}}, \bibinfo{author}{A.~D. {Myers}},
  \bibinfo{author}{N.~P. {Ross}}, \bibinfo{author}{D.~J. {Schlegel}},
  \bibinfo{author}{J.~F. {Hennawi}}, \bibinfo{author}{Y.~{Shen}},
  \bibinfo{author}{I.~{McGreer}}, \bibinfo{author}{M.~A. {Strauss}},
  \bibinfo{author}{A.~S. {Bolton}}, \bibinfo{author}{J.~{Bovy}}, \emph{et~al.},
  \bibinfo{journal}{\mnras} \bibinfo{volume}{\textbf{424}},
  \bibinfo{pages}{933} (\bibinfo{date}{Aug. 2012}), \eprint{1203.5306}.
\bibitem{2005ApJ...627L...1A}
\bibinfo{author}{K.~L. {Adelberger}} and \bibinfo{author}{C.~C. {Steidel}},
  \bibinfo{journal}{\apjl} \bibinfo{volume}{\textbf{627}}, \bibinfo{pages}{L1}
  (\bibinfo{date}{Jul. 2005}), \eprint{arXiv:astro-ph/0505546}.
\bibitem{2007ApJ...654..115C}
\bibinfo{author}{A.~L. {Coil}}, \bibinfo{author}{J.~F. {Hennawi}},
  \bibinfo{author}{J.~A. {Newman}}, \bibinfo{author}{M.~C. {Cooper}}, and
  \bibinfo{author}{M.~{Davis}}, \bibinfo{journal}{\apj}
  \bibinfo{volume}{\textbf{654}}, \bibinfo{pages}{115} (\bibinfo{date}{Jan.
  2007}).
\bibitem{2009MNRAS.397.1862P}
\bibinfo{author}{N.~{Padmanabhan}}, \bibinfo{author}{M.~{White}},
  \bibinfo{author}{P.~{Norberg}}, and \bibinfo{author}{C.~{Porciani}},
  \bibinfo{journal}{\mnras} \bibinfo{volume}{\textbf{397}},
  \bibinfo{pages}{1862} (\bibinfo{date}{Aug. 2009}), \eprint{0802.2105}.
\bibitem{Vikas2013}
\bibinfo{author}{Vikas++}, \bibinfo{title}{\emph{{QSO - CIV cross-correlation
  (should be done in a couple of weeks, update)}}}, , in preparation
  (\bibinfo{date}{2013}).
\bibitem{1982MNRAS.198...91C}
\bibinfo{author}{R.~F. {Carswell}}, \bibinfo{author}{J.~A.~J. {Whelan}},
  \bibinfo{author}{M.~G. {Smith}}, \bibinfo{author}{A.~{Boksenberg}}, and
  \bibinfo{author}{D.~{Tytler}}, \bibinfo{journal}{\mnras}
  \bibinfo{volume}{\textbf{198}}, \bibinfo{pages}{91} (\bibinfo{date}{Jan.
  1982}).
\bibitem{1986ApJ...309...19M}
\bibinfo{author}{H.~S. {Murdoch}}, \bibinfo{author}{R.~W. {Hunstead}},
  \bibinfo{author}{M.~{Pettini}}, and \bibinfo{author}{J.~C. {Blades}},
  \bibinfo{journal}{\apj} \bibinfo{volume}{\textbf{309}}, \bibinfo{pages}{19}
  (\bibinfo{date}{Oct. 1986}).
\bibitem{1988ApJ...327..570B}
\bibinfo{author}{S.~{Bajtlik}}, \bibinfo{author}{R.~C. {Duncan}}, and
  \bibinfo{author}{J.~P. {Ostriker}}, \bibinfo{journal}{\apj}
  \bibinfo{volume}{\textbf{327}}, \bibinfo{pages}{570} (\bibinfo{date}{Apr.
  1988}).
\bibitem{2000ApJS..130...67S}
\bibinfo{author}{J.~{Scott}}, \bibinfo{author}{J.~{Bechtold}},
  \bibinfo{author}{A.~{Dobrzycki}}, and \bibinfo{author}{V.~P. {Kulkarni}},
  \bibinfo{journal}{\apjs} \bibinfo{volume}{\textbf{130}}, \bibinfo{pages}{67}
  (\bibinfo{date}{Sep. 2000}), \eprint{arXiv:astro-ph/0004155}.
\bibitem{1986ApJ...303L..27J}
\bibinfo{author}{P.~{Jakobsen}}, \bibinfo{author}{M.~A.~C. {Perryman}},
  \bibinfo{author}{S.~{di Serego Alighieri}}, \bibinfo{author}{M.~H. {Ulrich}},
  and \bibinfo{author}{F.~{Macchetto}}, \bibinfo{journal}{\apjl}
  \bibinfo{volume}{\textbf{303}}, \bibinfo{pages}{L27} (\bibinfo{date}{Apr.
  1986}).
\bibitem{1989ApJ...336..550C}
\bibinfo{author}{A.~P.~S. {Crotts}}, \bibinfo{journal}{\apj}
  \bibinfo{volume}{\textbf{336}}, \bibinfo{pages}{550} (\bibinfo{date}{Jan.
  1989}).
\bibitem{1992A&A...258..234M}
\bibinfo{author}{P.~{Moller}} and \bibinfo{author}{P.~{Kjaergaard}},
  \bibinfo{journal}{\aap} \bibinfo{volume}{\textbf{258}}, \bibinfo{pages}{234}
  (\bibinfo{date}{May 1992}).
\bibitem{2001MNRAS.328..653L}
\bibinfo{author}{J.~{Liske}} and \bibinfo{author}{G.~M. {Williger}},
  \bibinfo{journal}{\mnras} \bibinfo{volume}{\textbf{328}},
  \bibinfo{pages}{653} (\bibinfo{date}{Dec. 2001}),
  \eprint{arXiv:astro-ph/0108239}.
\bibitem{1998ApJ...500..525S}
\bibinfo{author}{D.~J. {Schlegel}}, \bibinfo{author}{D.~P. {Finkbeiner}}, and
  \bibinfo{author}{M.~{Davis}}, \bibinfo{journal}{\apj}
  \bibinfo{volume}{\textbf{500}}, \bibinfo{pages}{525} (\bibinfo{date}{Jun.
  1998}).
\bibitem{2004ApJ...610..105S}
\bibinfo{author}{M.~{Schirber}}, \bibinfo{author}{J.~{Miralda-Escud{\'e}}}, and
  \bibinfo{author}{P.~{McDonald}}, \bibinfo{journal}{\apj}
  \bibinfo{volume}{\textbf{610}}, \bibinfo{pages}{105} (\bibinfo{date}{Jul.
  2004}), \eprint{arXiv:astro-ph/0307563}.
\bibitem{2004ApJ...610..642C}
\bibinfo{author}{R.~A.~C. {Croft}}, \bibinfo{journal}{\apj}
  \bibinfo{volume}{\textbf{610}}, \bibinfo{pages}{642} (\bibinfo{date}{Aug.
  2004}), \eprint{arXiv:astro-ph/0310890}.
\bibitem{2005MNRAS.361.1015R}
\bibinfo{author}{E.~{Rollinde}}, \bibinfo{author}{R.~{Srianand}},
  \bibinfo{author}{T.~{Theuns}}, \bibinfo{author}{P.~{Petitjean}}, and
  \bibinfo{author}{H.~{Chand}}, \bibinfo{journal}{\mnras}
  \bibinfo{volume}{\textbf{361}}, \bibinfo{pages}{1015} (\bibinfo{date}{Aug.
  2005}), \eprint{arXiv:astro-ph/0502284}.
\bibitem{2007MNRAS.377..657G}
\bibinfo{author}{R.~{Guimar{\~a}es}}, \bibinfo{author}{P.~{Petitjean}},
  \bibinfo{author}{E.~{Rollinde}}, \bibinfo{author}{R.~R. {de Carvalho}},
  \bibinfo{author}{S.~G. {Djorgovski}}, \bibinfo{author}{R.~{Srianand}},
  \bibinfo{author}{A.~{Aghaee}}, and \bibinfo{author}{S.~{Castro}},
  \bibinfo{journal}{\mnras} \bibinfo{volume}{\textbf{377}},
  \bibinfo{pages}{657} (\bibinfo{date}{May 2007}),
  \eprint{arXiv:astro-ph/0702369}.
\bibitem{2008MNRAS.387..377K}
\bibinfo{author}{Y.-R. {Kim}} and \bibinfo{author}{R.~A.~C. {Croft}},
  \bibinfo{journal}{\mnras} \bibinfo{volume}{\textbf{387}},
  \bibinfo{pages}{377} (\bibinfo{date}{Jun. 2008}).
\bibitem{2011AJ....142...72E}
\bibinfo{author}{D.~J. {Eisenstein}}, \bibinfo{author}{D.~H. {Weinberg}},
  \bibinfo{author}{E.~{Agol}}, \bibinfo{author}{H.~{Aihara}},
  \bibinfo{author}{C.~{Allende Prieto}}, \bibinfo{author}{S.~F. {Anderson}},
  \bibinfo{author}{J.~A. {Arns}}, \bibinfo{author}{{\'E}.~{Aubourg}},
  \bibinfo{author}{S.~{Bailey}}, \bibinfo{author}{E.~{Balbinot}},
  \emph{et~al.}, \bibinfo{journal}{\aj} \bibinfo{volume}{\textbf{142}},
  \bibinfo{pages}{72} (\bibinfo{date}{Sep. 2011}), \eprint{1101.1529}.
\bibitem{2013AJ....145...10D}
\bibinfo{author}{K.~S. {Dawson}}, \bibinfo{author}{D.~J. {Schlegel}},
  \bibinfo{author}{C.~P. {Ahn}}, \bibinfo{author}{S.~F. {Anderson}},
  \bibinfo{author}{{\'E}.~{Aubourg}}, \bibinfo{author}{S.~{Bailey}},
  \bibinfo{author}{R.~H. {Barkhouser}}, \bibinfo{author}{J.~E. {Bautista}},
  \bibinfo{author}{A.~{Beifiori}}, \bibinfo{author}{A.~A. {Berlind}},
  \emph{et~al.}, \bibinfo{journal}{\aj} \bibinfo{volume}{\textbf{145}},
  \bibinfo{pages}{10}, \bibinfo{eid}{10} (\bibinfo{date}{Jan. 2013}),
  \eprint{1208.0022}.
\bibitem{2012A&A...548A..66P}
\bibinfo{author}{I.~{P{\^a}ris}}, \bibinfo{author}{P.~{Petitjean}},
  \bibinfo{author}{{\'E}.~{Aubourg}}, \bibinfo{author}{S.~{Bailey}},
  \bibinfo{author}{N.~P. {Ross}}, \bibinfo{author}{A.~D. {Myers}},
  \bibinfo{author}{M.~A. {Strauss}}, \bibinfo{author}{S.~F. {Anderson}},
  \bibinfo{author}{E.~{Arnau}}, \bibinfo{author}{J.~{Bautista}}, \emph{et~al.},
  \bibinfo{journal}{\aap} \bibinfo{volume}{\textbf{548}}, \bibinfo{pages}{A66},
  \bibinfo{eid}{A66} (\bibinfo{date}{Dec. 2012}), \eprint{1210.5166}.
\bibitem{1987MNRAS.227....1K}
\bibinfo{author}{N.~{Kaiser}}, \bibinfo{journal}{\mnras}
  \bibinfo{volume}{\textbf{227}}, \bibinfo{pages}{1} (\bibinfo{date}{Jul.
  1987}).
\bibitem{HAMIL92}
\bibinfo{author}{A.~J.~S. {Hamilton}}, \bibinfo{journal}{\apjl}
  \bibinfo{volume}{\textbf{385}}, \bibinfo{pages}{L5} (\bibinfo{date}{Jan.
  1992}).
\bibitem{2011ApJS..192...18K}
\bibinfo{author}{E.~{Komatsu}}, \bibinfo{author}{K.~M. {Smith}},
  \bibinfo{author}{J.~{Dunkley}}, \bibinfo{author}{C.~L. {Bennett}},
  \bibinfo{author}{B.~{Gold}}, \bibinfo{author}{G.~{Hinshaw}},
  \bibinfo{author}{N.~{Jarosik}}, \bibinfo{author}{D.~{Larson}},
  \bibinfo{author}{M.~R. {Nolta}}, \bibinfo{author}{L.~{Page}}, \emph{et~al.},
  \bibinfo{journal}{\apjs} \bibinfo{volume}{\textbf{192}}, \bibinfo{pages}{18},
  \bibinfo{eid}{18} (\bibinfo{date}{Feb. 2011}), \eprint{1001.4538}.
\bibitem{2012ApJS..203...21A}
\bibinfo{author}{C.~P. {Ahn}}, \bibinfo{author}{R.~{Alexandroff}},
  \bibinfo{author}{C.~{Allende Prieto}}, \bibinfo{author}{S.~F. {Anderson}},
  \bibinfo{author}{T.~{Anderton}}, \bibinfo{author}{B.~H. {Andrews}},
  \bibinfo{author}{{\'E}.~{Aubourg}}, \bibinfo{author}{S.~{Bailey}},
  \bibinfo{author}{E.~{Balbinot}}, \bibinfo{author}{R.~{Barnes}},
  \emph{et~al.}, \bibinfo{journal}{\apjs} \bibinfo{volume}{\textbf{203}},
  \bibinfo{pages}{21}, \bibinfo{eid}{21} (\bibinfo{date}{Dec. 2012}),
  \eprint{1207.7137}.
\bibitem{2012AJ....144..144B}
\bibinfo{author}{A.~S. {Bolton}}, \bibinfo{author}{D.~J. {Schlegel}},
  \bibinfo{author}{{\'E}.~{Aubourg}}, \bibinfo{author}{S.~{Bailey}},
  \bibinfo{author}{V.~{Bhardwaj}}, \bibinfo{author}{J.~R. {Brownstein}},
  \bibinfo{author}{S.~{Burles}}, \bibinfo{author}{Y.-M. {Chen}},
  \bibinfo{author}{K.~{Dawson}}, \bibinfo{author}{D.~J. {Eisenstein}},
  \emph{et~al.}, \bibinfo{journal}{\aj} \bibinfo{volume}{\textbf{144}},
  \bibinfo{pages}{144}, \bibinfo{eid}{144} (\bibinfo{date}{Nov. 2012}),
  \eprint{1207.7326}.
\bibitem{1998AJ....116.3040G}
\bibinfo{author}{J.~E. {Gunn}}, \bibinfo{author}{M.~{Carr}},
  \bibinfo{author}{C.~{Rockosi}}, \bibinfo{author}{M.~{Sekiguchi}},
  \bibinfo{author}{K.~{Berry}}, \bibinfo{author}{B.~{Elms}},
  \bibinfo{author}{E.~{de Haas}}, \bibinfo{author}{{\v Z}.~{Ivezi{\' c}}},
  \bibinfo{author}{G.~{Knapp}}, \bibinfo{author}{R.~{Lupton}}, \emph{et~al.},
  \bibinfo{journal}{\aj} \bibinfo{volume}{\textbf{116}}, \bibinfo{pages}{3040}
  (\bibinfo{date}{Dec. 1998}).
\bibitem{1996AJ....111.1748F}
\bibinfo{author}{M.~{Fukugita}}, \bibinfo{author}{T.~{Ichikawa}},
  \bibinfo{author}{J.~E. {Gunn}}, \bibinfo{author}{M.~{Doi}},
  \bibinfo{author}{K.~{Shimasaku}}, and \bibinfo{author}{D.~P. {Schneider}},
  \bibinfo{journal}{\aj} \bibinfo{volume}{\textbf{111}}, \bibinfo{pages}{1748}
  (\bibinfo{date}{Apr. 1996}).
\bibitem{2006AJ....131.2332G}
\bibinfo{author}{J.~E. {Gunn}}, \bibinfo{author}{W.~A. {Siegmund}},
  \bibinfo{author}{E.~J. {Mannery}}, \bibinfo{author}{R.~E. {Owen}},
  \bibinfo{author}{C.~L. {Hull}}, \bibinfo{author}{R.~F. {Leger}},
  \bibinfo{author}{L.~N. {Carey}}, \bibinfo{author}{G.~R. {Knapp}},
  \bibinfo{author}{D.~G. {York}}, \bibinfo{author}{W.~N. {Boroski}},
  \emph{et~al.}, \bibinfo{journal}{\aj} \bibinfo{volume}{\textbf{131}},
  \bibinfo{pages}{2332} (\bibinfo{date}{Apr. 2006}),
  \eprint{arXiv:astro-ph/0602326}.
\bibitem{2012arXiv1208.2233S}
\bibinfo{author}{S.~{Smee}}, \bibinfo{author}{J.~E. {Gunn}},
  \bibinfo{author}{A.~{Uomoto}}, \bibinfo{author}{N.~{Roe}},
  \bibinfo{author}{D.~{Schlegel}}, \bibinfo{author}{C.~M. {Rockosi}},
  \bibinfo{author}{M.~A. {Carr}}, \bibinfo{author}{F.~{Leger}},
  \bibinfo{author}{K.~S. {Dawson}}, \bibinfo{author}{M.~D. {Olmstead}},
  \emph{et~al.}, \bibinfo{journal}{ArXiv e-prints}  (\bibinfo{date}{Aug.
  2012}), \eprint{1208.2233}.
\bibitem{2000AJ....120.1579Y}
\bibinfo{author}{D.~G. {York}}, \bibinfo{author}{J.~{Adelman}},
  \bibinfo{author}{J.~E. {Anderson}}, \bibinfo{author}{S.~F. {Anderson}},
  \bibinfo{author}{J.~{Annis}}, \bibinfo{author}{N.~A. {Bahcall}},
  \bibinfo{author}{J.~A. {Bakken}}, \bibinfo{author}{R.~{Barkhouser}},
  \bibinfo{author}{S.~{Bastian}}, \bibinfo{author}{E.~{Berman}}, \emph{et~al.},
  \bibinfo{journal}{\aj} \bibinfo{volume}{\textbf{120}}, \bibinfo{pages}{1579}
  (\bibinfo{date}{Sep. 2000}).
\bibitem{2013AJ....145...69L}
\bibinfo{author}{K.-G. {Lee}}, \bibinfo{author}{S.~{Bailey}},
  \bibinfo{author}{L.~E. {Bartsch}}, \bibinfo{author}{W.~{Carithers}},
  \bibinfo{author}{K.~S. {Dawson}}, \bibinfo{author}{D.~{Kirkby}},
  \bibinfo{author}{B.~{Lundgren}}, \bibinfo{author}{D.~{Margala}},
  \bibinfo{author}{N.~{Palanque-Delabrouille}}, \bibinfo{author}{M.~M.
  {Pieri}}, \emph{et~al.}, \bibinfo{journal}{\aj}
  \bibinfo{volume}{\textbf{145}}, \bibinfo{pages}{69}, \bibinfo{eid}{69}
  (\bibinfo{date}{Mar. 2013}), \eprint{1211.5146}.
\bibitem{2012ApJS..199....3R}
\bibinfo{author}{N.~P. {Ross}}, \bibinfo{author}{A.~D. {Myers}},
  \bibinfo{author}{E.~S. {Sheldon}}, \bibinfo{author}{C.~{Y{\`e}che}},
  \bibinfo{author}{M.~A. {Strauss}}, \bibinfo{author}{J.~{Bovy}},
  \bibinfo{author}{J.~A. {Kirkpatrick}}, \bibinfo{author}{G.~T. {Richards}},
  \bibinfo{author}{{\'E}.~{Aubourg}}, \bibinfo{author}{M.~R. {Blanton}},
  \emph{et~al.}, \bibinfo{journal}{\apjs} \bibinfo{volume}{\textbf{199}},
  \bibinfo{pages}{3}, \bibinfo{eid}{3} (\bibinfo{date}{Mar. 2012}),
  \eprint{1105.0606}.
\bibitem{2010A&A...523A..14Y}
\bibinfo{author}{C.~{Y{\`e}che}}, \bibinfo{author}{P.~{Petitjean}},
  \bibinfo{author}{J.~{Rich}}, \bibinfo{author}{E.~{Aubourg}},
  \bibinfo{author}{N.~{Busca}}, \bibinfo{author}{J.-C. {Hamilton}},
  \bibinfo{author}{J.-M. {Le Goff}}, \bibinfo{author}{I.~{Paris}},
  \bibinfo{author}{S.~{Peirani}}, \bibinfo{author}{C.~{Pichon}}, \emph{et~al.},
  \bibinfo{journal}{\aap} \bibinfo{volume}{\textbf{523}}, \bibinfo{pages}{A14},
  \bibinfo{eid}{A14} (\bibinfo{date}{Nov. 2010}).
\bibitem{2011ApJ...743..125K}
\bibinfo{author}{J.~A. {Kirkpatrick}}, \bibinfo{author}{D.~J. {Schlegel}},
  \bibinfo{author}{N.~P. {Ross}}, \bibinfo{author}{A.~D. {Myers}},
  \bibinfo{author}{J.~F. {Hennawi}}, \bibinfo{author}{E.~S. {Sheldon}},
  \bibinfo{author}{D.~P. {Schneider}}, and \bibinfo{author}{B.~A. {Weaver}},
  \bibinfo{journal}{\apj} \bibinfo{volume}{\textbf{743}}, \bibinfo{pages}{125},
  \bibinfo{eid}{125} (\bibinfo{date}{Dec. 2011}), \eprint{1104.4995}.
\bibitem{2011ApJ...729..141B}
\bibinfo{author}{J.~{Bovy}}, \bibinfo{author}{J.~F. {Hennawi}},
  \bibinfo{author}{D.~W. {Hogg}}, \bibinfo{author}{A.~D. {Myers}},
  \bibinfo{author}{J.~A. {Kirkpatrick}}, \bibinfo{author}{D.~J. {Schlegel}},
  \bibinfo{author}{N.~P. {Ross}}, \bibinfo{author}{E.~S. {Sheldon}},
  \bibinfo{author}{I.~D. {McGreer}}, \bibinfo{author}{D.~P. {Schneider}},
  \emph{et~al.}, \bibinfo{journal}{\apj} \bibinfo{volume}{\textbf{729}},
  \bibinfo{pages}{141}, \bibinfo{eid}{141} (\bibinfo{date}{Mar. 2011}),
  \eprint{1011.6392}.
\bibitem{2012JCAP...11..059F}
\bibinfo{author}{A.~{Font-Ribera}}, \bibinfo{author}{J.~{Miralda-Escud{\'e}}},
  \bibinfo{author}{E.~{Arnau}}, \bibinfo{author}{B.~{Carithers}},
  \bibinfo{author}{K.-G. {Lee}}, \bibinfo{author}{P.~{Noterdaeme}},
  \bibinfo{author}{I.~{P{\^a}ris}}, \bibinfo{author}{P.~{Petitjean}},
  \bibinfo{author}{J.~{Rich}}, \bibinfo{author}{E.~{Rollinde}}, \emph{et~al.},
  \bibinfo{journal}{\jcap} \bibinfo{volume}{\textbf{11}}, \bibinfo{pages}{59},
  \bibinfo{eid}{059} (\bibinfo{date}{Nov. 2012}), \eprint{1209.4596}.
\bibitem{2012A&A...547L...1N}
\bibinfo{author}{P.~{Noterdaeme}}, \bibinfo{author}{P.~{Petitjean}},
  \bibinfo{author}{W.~C. {Carithers}}, \bibinfo{author}{I.~{P{\^a}ris}},
  \bibinfo{author}{A.~{Font-Ribera}}, \bibinfo{author}{S.~{Bailey}},
  \bibinfo{author}{E.~{Aubourg}}, \bibinfo{author}{D.~{Bizyaev}},
  \bibinfo{author}{G.~{Ebelke}}, \bibinfo{author}{H.~{Finley}}, \emph{et~al.},
  \bibinfo{journal}{\aap} \bibinfo{volume}{\textbf{547}}, \bibinfo{pages}{L1},
  \bibinfo{eid}{L1} (\bibinfo{date}{Nov. 2012}), \eprint{1210.1213}.
\bibitem{Carithers2013}
\bibinfo{author}{Carithers++}, \bibinfo{title}{\emph{{Concordance DLA
  Catalogue}}}, , in preparation (\bibinfo{date}{2013}).
\bibitem{2012AJ....143...51L}
\bibinfo{author}{K.-G. {Lee}}, \bibinfo{author}{N.~{Suzuki}}, and
  \bibinfo{author}{D.~N. {Spergel}}, \bibinfo{journal}{\aj}
  \bibinfo{volume}{\textbf{143}}, \bibinfo{pages}{51}, \bibinfo{eid}{51}
  (\bibinfo{date}{Feb. 2012}).
\bibitem{2008ApJ...681..831F}
\bibinfo{author}{C.-A. {Faucher-Gigu{\`e}re}}, \bibinfo{author}{J.~X.
  {Prochaska}}, \bibinfo{author}{A.~{Lidz}}, \bibinfo{author}{L.~{Hernquist}},
  and \bibinfo{author}{M.~{Zaldarriaga}}, \bibinfo{journal}{\apj}
  \bibinfo{volume}{\textbf{681}}, \bibinfo{pages}{831} (\bibinfo{date}{Jul.
  2008}), \eprint{0709.2382}.
\bibitem{2013A&A...552A..96B}
\bibinfo{author}{N.~G. {Busca}}, \bibinfo{author}{T.~{Delubac}},
  \bibinfo{author}{J.~{Rich}}, \bibinfo{author}{S.~{Bailey}},
  \bibinfo{author}{A.~{Font-Ribera}}, \bibinfo{author}{D.~{Kirkby}},
  \bibinfo{author}{J.-M. {Le Goff}}, \bibinfo{author}{M.~M. {Pieri}},
  \bibinfo{author}{A.~{Slosar}}, \bibinfo{author}{{\'E}.~{Aubourg}},
  \emph{et~al.}, \bibinfo{journal}{\aap} \bibinfo{volume}{\textbf{552}},
  \bibinfo{pages}{A96}, \bibinfo{eid}{A96} (\bibinfo{date}{Apr. 2013}),
  \eprint{1211.2616}.
\bibitem{2010MNRAS.405.2302H}
\bibinfo{author}{P.~C. {Hewett}} and \bibinfo{author}{V.~{Wild}},
  \bibinfo{journal}{\mnras} \bibinfo{volume}{\textbf{405}},
  \bibinfo{pages}{2302} (\bibinfo{date}{Jul. 2010}), \eprint{1003.3017}.
\bibitem{2001AJ....122..549V}
\bibinfo{author}{D.~E. {Vanden Berk}}, \bibinfo{author}{G.~T. {Richards}},
  \bibinfo{author}{A.~{Bauer}}, \bibinfo{author}{M.~A. {Strauss}},
  \bibinfo{author}{D.~P. {Schneider}}, \bibinfo{author}{T.~M. {Heckman}},
  \bibinfo{author}{D.~G. {York}}, \bibinfo{author}{P.~B. {Hall}},
  \bibinfo{author}{X.~{Fan}}, \bibinfo{author}{G.~R. {Knapp}}, \emph{et~al.},
  \bibinfo{journal}{\aj} \bibinfo{volume}{\textbf{122}}, \bibinfo{pages}{549}
  (\bibinfo{date}{Aug. 2001}), \eprint{arXiv:astro-ph/0105231}.
\bibitem{2012ApJ...752..162S}
\bibinfo{author}{J.~M. {Shull}}, \bibinfo{author}{M.~{Stevans}}, and
  \bibinfo{author}{C.~W. {Danforth}}, \bibinfo{journal}{\apj}
  \bibinfo{volume}{\textbf{752}}, \bibinfo{pages}{162}, \bibinfo{eid}{162}
  (\bibinfo{date}{Jun. 2012}), \eprint{1204.3908}.
\bibitem{1994MNRAS.266..343M}
\bibinfo{author}{J.~{Miralda-Escud{\'e}}} and \bibinfo{author}{M.~J. {Rees}},
  \bibinfo{journal}{\mnras} \bibinfo{volume}{\textbf{266}},
  \bibinfo{pages}{343} (\bibinfo{date}{Jan. 1994}).
\bibitem{2009ApJ...694..842M}
\bibinfo{author}{M.~{McQuinn}}, \bibinfo{author}{A.~{Lidz}},
  \bibinfo{author}{M.~{Zaldarriaga}}, \bibinfo{author}{L.~{Hernquist}},
  \bibinfo{author}{P.~F. {Hopkins}}, \bibinfo{author}{S.~{Dutta}}, and
  \bibinfo{author}{C.-A. {Faucher-Gigu{\`e}re}}, \bibinfo{journal}{\apj}
  \bibinfo{volume}{\textbf{694}}, \bibinfo{pages}{842} (\bibinfo{date}{Apr.
  2009}), \eprint{0807.2799}.
\bibitem{2003ApJ...585...34M}
\bibinfo{author}{P.~{McDonald}}, \bibinfo{journal}{\apj}
  \bibinfo{volume}{\textbf{585}}, \bibinfo{pages}{34} (\bibinfo{date}{Mar.
  2003}), \eprint{arXiv:astro-ph/0108064}.
\bibitem{2012JCAP...07..028F}
\bibinfo{author}{A.~{Font-Ribera}} and
  \bibinfo{author}{J.~{Miralda-Escud{\'e}}}, \bibinfo{journal}{\jcap}
  \bibinfo{volume}{\textbf{7}}, \bibinfo{pages}{28} (\bibinfo{date}{Jul.
  2012}), \eprint{1205.2018}.
\bibitem{2013JCAP...04..026S}
\bibinfo{author}{A.~{Slosar}}, \bibinfo{author}{V.~{Ir{\v s}i{\v c}}},
  \bibinfo{author}{D.~{Kirkby}}, \bibinfo{author}{S.~{Bailey}},
  \bibinfo{author}{N.~G. {Busca}}, \bibinfo{author}{T.~{Delubac}},
  \bibinfo{author}{J.~{Rich}}, \bibinfo{author}{{\'E}.~{Aubourg}},
  \bibinfo{author}{J.~E. {Bautista}}, \bibinfo{author}{V.~{Bhardwaj}},
  \emph{et~al.}, \bibinfo{journal}{\jcap} \bibinfo{volume}{\textbf{4}},
  \bibinfo{pages}{26}, \bibinfo{eid}{026} (\bibinfo{date}{Apr. 2013}),
  \eprint{1301.3459}.

\end{thebibliography}
\bibliographystyle{revtex}



\end{document}